\documentclass[11pt]{article} %
\usepackage{etoolbox}
\newtoggle{long}
\toggletrue{long} %

\newcommand{\longversion}[1]{\iftoggle{long}{#1}{}}
\newcommand{\shortversion}[1]{\iftoggle{long}{}{#1}}

\shortversion{\usepackage[margin=1.0in]{geometry}}
\longversion{
\usepackage[margin=1.3in]{geometry}
}
\usepackage[utf8]{inputenc}
\usepackage[T1]{fontenc}
\longversion{\usepackage[USenglish]{babel}}
\usepackage{lmodern} %
\usepackage{graphicx}
\usepackage{subcaption} \DeclareCaptionSubType[roman]{figure} %
\usepackage{enumitem} \setlist[enumerate,1]{label={(\roman*)}} \shortversion{\setlist{noitemsep,topsep=0pt,parsep=0pt,partopsep=0pt}}%
\usepackage{microtype}

\newcommand{\titel}{Dynamics of Cycles in Polyhedra I:\protect\\ The Isolation Lemma} %
\graphicspath{{./figures/}{../figures/}}

\usepackage{algorithm} %
\usepackage[noend]{algpseudocode} %

\usepackage[usenames]{xcolor} %
\definecolor{hellblau}{rgb}{0.2,0.4,1} %
\definecolor{dunkelblau}{rgb}{0,0,0.8}
\definecolor{dunkelgruen}{rgb}{0,0.5,0}
\usepackage[
	pdftex,
	colorlinks,
	linkcolor=dunkelblau,
	urlcolor=dunkelblau,
	citecolor=dunkelgruen,
	bookmarks=true,
	linktocpage=true,
	pdfauthor={},
	pdfsubject={}%
]{hyperref}
\usepackage{pdfpages} %

\makeatletter
\def\blfootnote{\xdef\@thefnmark{}\@footnotetext}
\makeatother

\usepackage{amsmath} %
\usepackage{amsthm} 
\usepackage{amsfonts} %
\theoremstyle{plain} %
	\newtheorem{satz}{Satz}[] %
	\newtheorem{theorem}[satz]{Theorem}
	\newtheorem*{theorem*}{Theorem}
	\newtheorem{lemma}[satz]{Lemma}
	\newtheorem*{lemma*}{Lemma}
	\newtheorem{corollary}[satz]{Corollary}

\theoremstyle{remark} %
\theoremstyle{definition} %
	\newtheorem{definition}[satz]{Definition}

\newcommand{\pulls}[3]{$#1 {\xleftarrow{#2}} #3$}

	\title{\titel
	}
	\author{Jan Kessler\\Institute of Mathematics\\University of Cologne, Germany
		\and Jens M.\ Schmidt\thanks{This research is supported by the grant SCHM 3186/2-1 (401348462) from the Deutsche Forschungsgemeinschaft (DFG, German Research Foundation), and within the project~57447800 by DAAD (as part of BMBF, Germany) and the Ministry of Education Science, Research and Sport of the Slovak Republic. A part of this research was carried out at TU Ilmenau.}%
					\\Institute of Mathematics\\University of Cologne, Germany}
	\date{}

\begin{document}
\maketitle

\begin{abstract}
A cycle $C$ of a graph $G$ is \emph{isolating} if every component of $G-V(C)$ is a single vertex. We show that isolating cycles in polyhedral graphs can be extended to larger ones: every isolating cycle $C$ of length $6 \leq |E(C)| < \left \lfloor \frac{2}{3}(|V(G)|+4) \right \rfloor$ implies an isolating cycle $C'$ of larger length that contains $V(C)$. By ``hopping'' iteratively to such larger cycles, we obtain a powerful and very general inductive motor for proving long cycles and computing them (we will give an algorithm with quadratic running time). This is the first step towards the so far elusive quest of finding a universal induction that captures longest cycles of polyhedral graph classes.

Our motor provides also a method to prove linear lower bounds on the length of Tutte cycles, as $C'$ will be a Tutte cycle of $G$ if $C$ is. We prove in addition that $|E(C')| \leq |E(C)|+3$ if $G$ contains no face of size five, which gives a new tool for results about cycle spectra, and provides evidence that faces of size five may obstruct long cycles in many graph classes. As a sample application, we test our motor on the following so far unsettled conjecture about essentially 4-connected graphs.

A planar graph is \emph{essentially $4$-connected} if it is 3-connected and every of its 3-separators is the neighborhood of a single vertex. Essentially $4$-connected graphs have been thoroughly investigated throughout literature as the subject of Hamiltonicity studies. Jackson and Wormald proved that every essentially 4-connected planar graph $G$ on $n$ vertices contains a cycle of length at least $\frac{2}{5}(n+2)$, and this result has recently been improved multiple times, culminating in the lower bound $\frac{5}{8}(n+2)$. However, the currently best known upper bound is given by an infinite family of such graphs in which no graph $G$ contains a cycle that is longer than $\left \lfloor \frac{2}{3}(n+4) \right \rfloor$; this upper bound is still unmatched.

Using isolating cycles, we improve the lower bound to match the upper. This settles the long-standing open problem of determining the circumference of essentially 4-connected planar graphs. All our results are tight.
\end{abstract}

\section{Introduction}
One of the unchallenged milestones in graph theory is the result by Tutte~\cite{Tutte1956} in 1956 that every 4-connected planar graph is Hamiltonian. However, decreasing the connectedness assumption from~4 to~3 reveals infinitely many planar graphs that do not have long cycles: In fact, Moon and Moser~\cite{Moon1963} showed that there are infinitely many 3-connected planar (i.e.\ \emph{polyhedral}) graphs that have circumference at most $9n^{\log_32}$ for $n := |V(G)|$, and this upper bound is best possible up to constant factors, as there is a constant $c > 0$ such that every polyhedral graph contains a cycle of length at least $cn^{\log_32}$.

One of the biggest remaining open problems in this area ever since is to characterize properties \emph{between} connectivity~3 and~4 that imply long cycles (see Grünbaum and Walther~\cite[Theorem~1]{Gruenbaum1973} for a classic reference that summarizes an abundance of such results for polyhedral subclasses). Essential $4$-connectedness is such a property and will be a focus of this paper.

Indeed, essentially $4$-connected graphs have been thoroughly investigated throughout literature for this purpose. An upper bound for the circumference of essentially 4-connected planar graphs was given in~\cite{Fabrici2016} by an infinite family of such graphs on $n \geq 14$ vertices in which every graph $G$ satisfies $circ(G)=\lfloor \frac{2}{3}(n+4) \rfloor$; the graphs in this family are in addition maximal planar.
Regarding lower bounds, Jackson and Wormald~\cite{Jackson1992} proved in~1992 that $circ(G) \geq \frac{2}{5}(n+2)$ for every essentially 4-connected planar graph $G$ on $n$ vertices. Fabrici, Harant and Jendroľ~\cite{Fabrici2016} improved this lower bound to $circ(G) \ge \frac{1}{2}(n+4)$; this result in turn was recently strengthened to $circ(G) \geq \frac{3}{5}(n+2)$~\cite{Fabrici2020}, and then further to $circ(G) \geq \frac{5}{8}(n+2)$~\cite{Fabrici2020a}. For the restricted case of maximal planar essentially 4-connected graphs, the matching lower bound $circ(G) \geq \frac{2}{3}(n+4)$ was proven in~\cite{Fabrici2020b}; however, the methods used there are specific to maximal planar graphs. For the general polyhedral case, it is still an open conjecture that every essentially $4$-connected planar graph $G$ on $n$ vertices satisfies $circ(G) \geq \lfloor \frac{2}{3}(n+4) \rfloor$; while this conjecture has been an active research topic at workshops (such as the ILKE Workshops on Graph Theory) for over a decade\longversion{\footnote{personal communication with Jochen Harant}}, it was only recently explicitly stated in~\cite[Conjecture~2]{Fabrici2020b}.

Here, we show that $circ(G) \geq \lfloor \frac{2}{3}(n+4) \rfloor$ for every essentially 4-connected planar graph. This matches the upper bound given above tightly. In fact, we give a much more general result of which the previous statement is just an implication: Instead of essentially 4-connected polyhedral graphs, the result holds for all polyhedral graphs that contain an isolating cycle, and instead of just proving high circumference, we prove many different long cycle lengths as follows. Let a cycle $C$ of a graph $G$ be \emph{extendable} if $G$ contains a larger isolating cycle $C'$ such that $V(C) \subset V(C')$ and $|V(C')| \leq |V(C)| + 3+n_5(G)$, where $n_5(G)$ is the number of faces of size five in $G$. The following is our main result.

\begin{lemma}[Isolation Lemma]\label{lem:main}
Every isolating cycle of length $c < \min\{\lfloor \frac{2}{3}(n+4) \rfloor,n\}$ in a polyhedral graph $G$ on $n$ vertices is extendable.
\end{lemma}

We note that the assumption $c < n$ may equivalently be replaced with $n \geq 6$, as we have $\lfloor \frac{2}{3}(n+4) \rfloor \leq n$ if and only if $n \geq 6$.
While one part of the proof scheme for Lemma~\ref{lem:main} follows the established approach of using Tutte cycles in combination with the discharging method, we contribute an intricate intersection argument on the weight distribution between groups of neighboring faces, which we call tunnels. This method differs substantially from the ones used in~\cite{Fabrici2016,Fabrici2020b,Fabrici2020,Fabrici2020a} (for example, \cite{Fabrici2020b} exploits the inherent structure of maximal planar graphs) and is able to harness the dynamics of extending cycles in general polyhedral graphs. In particular, we will discharge weights along an unbounded number of faces, which was an obstacle that was not needed to overcome for the previous bounds.

The Isolation Lemma may be seen as a polyhedral relative of Woodall's Hopping Lemma~\cite{Woodall1973} that allows cycle extensions through common neighbors of cycle vertex pairs even when \emph{none} of these pairs have distance two in $C$. Despite this relation, the Isolation Lemma makes inherently use of planarity; in fact, it fails hard for non-planar graphs, as the graphs $K_{c,n-c}$ for any $c \geq 3$ \shortversion{show.}\longversion{ and $n > 2c$ and any (isolating) cycle of length $2c$ in these show. For $c = 2$, these graphs show also that 3-connectedness is necessary for planar graphs.}

We state some immediate corollaries of the Isolation Lemma.

\begin{corollary}\label{cor:manyCycleLengths}
Let $G$ be a polyhedral graph on $n \geq 6$ vertices with $n_5$ faces of size five. If $G$ contains an isolating cycle $C$, $G$ contains isolating cycles of at least $(\lfloor \frac{2}{3}(n+4) \rfloor-|E(C)|+1)/(3+n_5)$ different lengths in $\{|E(C)|,\dots,\lfloor \frac{2}{3}(n+4) \rfloor\}$, all of which contain $V(C)$.
\end{corollary}

In particular, for bipartite graphs, Corollary~\ref{cor:manyCycleLengths} implies the following.

\begin{corollary}\label{cor:bipartite}
If a bipartite polyhedral graph on $n \geq 6$ vertices contains an isolating cycle $C$, it contains an isolating cycle of every even length $l \in \{|E(C)|,\dots,\lfloor \frac{2}{3}(n+4) \rfloor\}$.
\end{corollary}

In other words, bipartite polyhedral graphs with an isolating cycle are bipancyclic in the given range. In view of the sheer number of results in Hamiltonicity studies that use subgraphs involving faces of size five (confer the Tutte Fragment or the fragment of Faulkner and Younger for classic examples), Lemma~\ref{lem:main} provides evidence that these faces are indeed key to a small circumference. Another corollary is that polyhedral graphs on $n$ vertices, in which all cycles have length less than $\min\{\lfloor \frac{2}{3}(n+4) \rfloor,n\}$, do not contain any isolating cycle; for example, this holds for the sufficiently large Moon-Moser graphs~\cite{Moon1963} and for every of the~18 graph classes of~\cite[Theorem~1]{Gruenbaum1973} that have shortness exponent less than~1.

Finally, the Isolation Lemma implies also the following theorem.

\begin{theorem}\label{thm:essential}
Every essentially 4-connected planar graph $G$ on $n$ vertices contains an isolating Tutte cycle of length at least $\min\{\lfloor \frac{2}{3}(n+4) \rfloor,n\}$.
\end{theorem}
\longversion{
\begin{proof}
It is well-known that every 3-connected plane graph on at most~10 vertices is Hamiltonian~\cite{Dillencourt1996}; these graphs contain in particular the essentially 4-connected planar graphs. Since every Hamiltonian cycle is isolating, this implies the theorem for every $n \leq 10$. We therefore assume $n \geq 11$; in particular, $\lfloor \frac{2}{3}(n+4) \rfloor \leq n$.
For $n \geq 11$, it was shown in~\cite[Lemma~4(i)]{Fabrici2016} that $G$ contains an isolating Tutte cycle $C$. %
Applying iteratively the Isolation Lemma to $C$ gives the claim and preserves a Tutte cycle, as no vertex of $C$ is deleted.
\end{proof}
}

Theorem~\ref{thm:essential} encompasses and strengthens most of the results known for the circumference of essentially 4-connected planar graphs, some of which can be found in~\cite{Fabrici2016,Grunbaum1976,Zhang1987}. At the same time and of independent interest, Theorem~\ref{thm:essential} allows to extend every isolating Tutte cycle $C$ of a polyhedral graph $G$ to an isolating Tutte cycle of $G$ of linear length in $|V(G)|$. We sketch algorithms computing such cycles efficiently at the end of this paper.

\longversion{\section{Preliminaries}
We use standard graph-theoretic terminology and consider only graphs that are finite, simple and undirected. For a vertex $v$ of a graph $G$, denote by $\deg_G(v)$ the degree of $v$ in $G$.
We omit subscripts if the graph $G$ is clear from the context. Two edges $e$ and $f$ are \emph{adjacent} if they share at least one end vertex. The \emph{distance} of two edges in a connected graph is the length of a shortest path that contains both. We denote a path of $G$ that visits the vertices $v_1,v_2,\dots,v_i$ in the given order by $v_1v_2 \dots v_i$.

}
A \emph{separator $S$} of a graph $G$ is a subset of $V(G)$ such that $G-S$ is disconnected; we call $S$ a $k$-\emph{separator} if $|S|=k$.
Let a cycle $C$ of a graph $G$ be \emph{isolating} if every component of $G-V(C)$ consists of a single vertex (see Figure~\ref{fig:IsolatingCycle1} for an example). We do not require that these single vertices have degree three (this differs e.g.\ from~\cite{Fabrici2016,Fabrici2020,Fabrici2020a}). A \emph{chord} of a cycle $C$ is an edge $vw \notin E(C)$ for which $v$ and $w$ are in $C$. 
By a result of Whitney~\cite{Whitney1932}, every 3-connected planar graph has a unique embedding into the plane (up to flipping and the choice of the outer face). Hence, we assume in the following that such graphs are equipped with a fixed planar embedding, i.e.\ are \emph{plane}, so that vertices are points and edges are point sets. Let $F(G)$ be the set of faces of a plane graph $G$.

\section{Proof of the Isolation Lemma}\label{sec:lemma}
Let $G = (V,E)$ be a 3-connected plane graph on $n$ vertices, and let $C$ be an isolating cycle of $G$ of length $c < \min\{\lfloor \frac{2}{3}(n+4) \rfloor,n\}$. We assume to the contrary that $C$ is not extendable. Then $c < \frac{2}{3}(n+3)$, as $c \geq \frac{2}{3}(n+3)$ implies $c \geq \lceil \frac{2}{3}(n+3) \rceil = \lfloor \frac{2}{3}(n+4) \rfloor$ for the integers $c$ and $n$, which is a contradiction.

\begin{figure}[!ht]
	\centering
	\subcaptionbox{An isolating cycle $C$ (fat edges) of an essentially 4-connected plane graph $G$; vertices in $V^+$ are not drawn. Here, all vertices of $V^- = \{a,b,d,e,g,h\}$ have degree three in $G$, and $H^-$ has no minor 1-face but would have one after contracting $yz$. The dashed chord $vw$ of $C$ is in $G$ but not in $H$, so that $f$ is a (thick major 1)-face of $H$ but not a face of $G$. The minor face $f_8$ has exactly the two arches $pgs$ and $pr$ (note that $ps$ is not an arch).
			\label{fig:IsolatingCycle1}}[0.51\linewidth]
			{\includegraphics[page=1,scale=0.6]{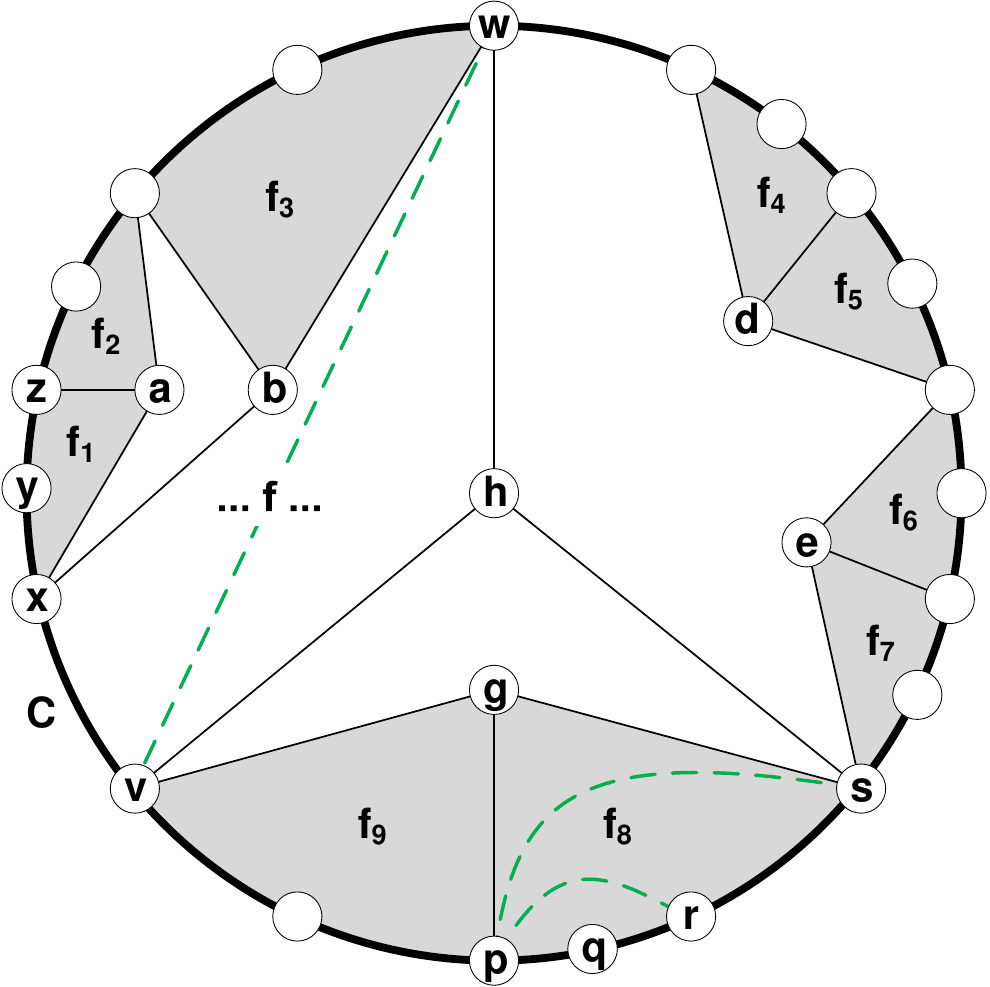}}
	\hspace{0.3cm}
	\subcaptionbox{The subgraph $H^-$ of $G$ (solid edges) and a tree $T^-$ constructed from $H^-$ (dashed edges). There are $|M^-| = 9 \geq |V^-| + 2 = 8$ minor faces in $H^-$ (depicted in grey), each of which is a thick 2- or 3-face that corresponds to a leaf of $T^-$.
			\label{fig:IsolatingCycle2}}%
			{\includegraphics[page=1,scale=0.6]{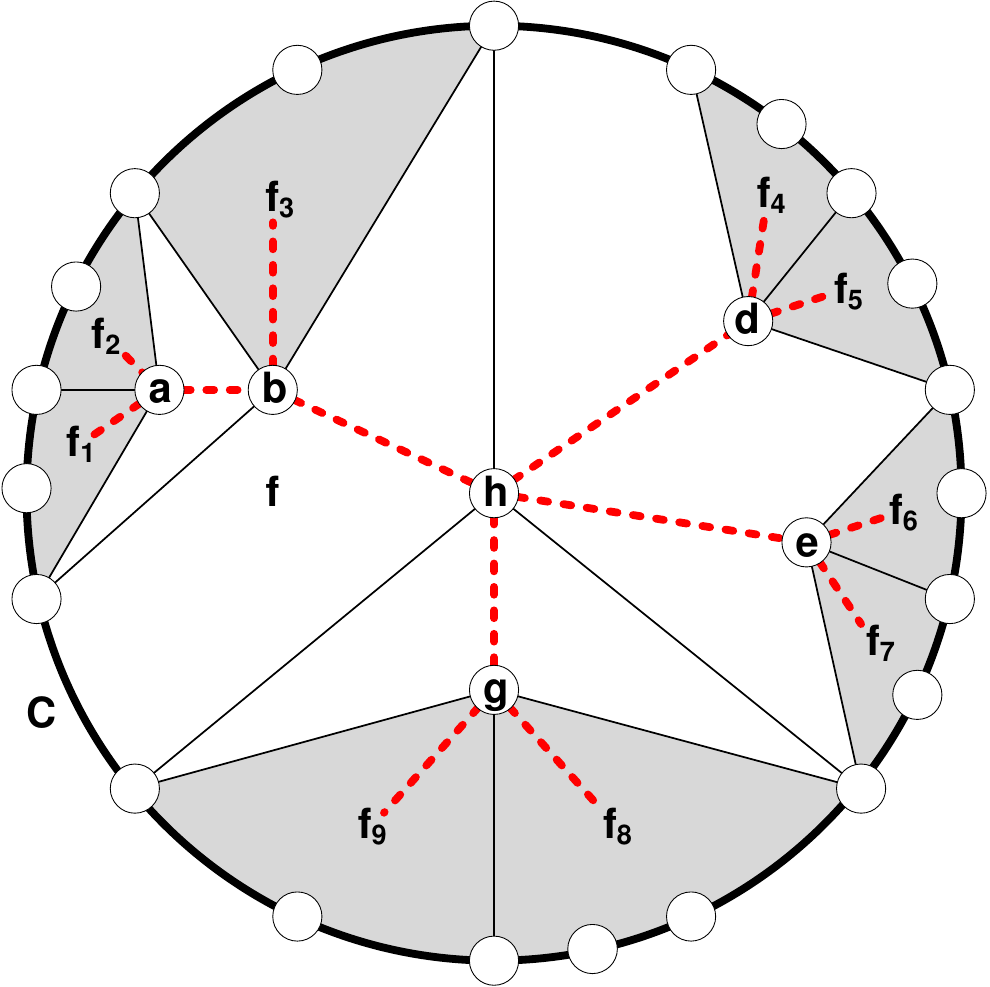}}
	\caption{}\label{fig:IsolatingCycle}
\end{figure}

Let $V^-$ be the subset of $V$ that is contained in one of the two \emph{regions} (i.e.\ maximal path-connected open sets) of $\mathbb{R}^2-C$, and let $V^+ := V-V(C)-V^-$. Without loss of generality, we assume $|V^-| \leq |V^+|$.
Since $c < n$ and $|V^-| \leq |V^+|$, we have $V^+ \neq \emptyset$. Let $H$ be the plane graph obtained from $G$ by either deleting all chords of $C$ if $V^- \neq \emptyset$ or otherwise deleting all chords of $C$ whose interior point sets are contained in the same region of $\mathbb{R}^2-C$ as $V^+$ (see Figure~\ref{fig:IsolatingCycle}). Let $H^- := H - V^+$ and $H^+ := H - V^- - (E(H^-) - E(C)))$.

For a face $f$, the edges of $C$ that are incident with $f$ are called $C$-\emph{edges} of $f$ and their number is denoted by $m_f$. A face $f$ is called $j$-\emph{face} if $m_f=j$. A $C$-\emph{vertex} of $f$ is a vertex that is incident to a $C$-edge of $f$. A $C$-edge of $f$ is \emph{extremal} if it is adjacent to at most one $C$-edge of $f$, and \emph{non-extremal} otherwise; a $C$-vertex of $f$ is \emph{extremal} if it is incident to at most one $C$-edge of $f$, and \emph{non-extremal} otherwise. If $f$ has an odd number of $C$-vertices or $C$-edges, we call its unique $C$-vertex or $C$-edge in the middle the \emph{middle} $C$-vertex or $C$-edge of $f$. Let two faces $f$ and $g$ of $H$ be \emph{opposite} if $f$ and $g$ have a common $C$-edge. If $f$ has a $C$-edge $e$, let the $e$-\emph{opposite face} of $f$ be the face of $H$ that is different from $f$ and incident to $e$.

Let a face $f$ of $H$ be \emph{thin} if $V^- = \emptyset$ and $f \in F(H^-)$, and \emph{thick} otherwise.
A face $f$ of $H$ is called \emph{minor} if it is either thin and incident to exactly one edge that is not in $C$ or thick and incident to exactly one vertex of $V^- \cup V^+$; otherwise, $f$ is called \emph{major} (this definition differs from~\cite{Fabrici2016,Fabrici2020,Fabrici2020a}).
Our motivation for minor faces stems from the fact that all their $C$-vertices are consecutive in $C$, which we will later use to impose structural restrictions on the incident edges to these vertices. Let $M^-$ and $M^+$ be the sets of minor faces of $H^-$ and $H^+$, respectively. For a minor thick face $f$ of $H$, let $v_f$ be the unique vertex of $V^+ \cup V^-$ that is incident to $f$.

We now show that the minor faces of $H$ are precisely the leaf faces of two suitably defined trees $T^-$ and $T^+$ that we construct from $H^-$ and $H^+$, respectively. If $V^- = \emptyset$, let $T^-$ be the weak dual of $H^-$. If $V^- \neq \emptyset$, let $T^-$ (we define $T^+$ analogously from $H^+$) be the graph with vertex set $M^- \cup V^-$ and the following edge-set (see Figure~\ref{fig:IsolatingCycle2}). First, for every face $f \in M^-$, add the edge $fv_f$ to $T^-$. Second, for every major face $f$ of $H^-$ (in arbitrary order), fix any vertex $v \in V^-$ that is incident to $f$ and add the edge $vw$ to $T^-$ for every vertex $w \in V^- - \{v\}$ that is incident to $f$. We prove that $T^-$ and $T^+$ are trees with the following properties.

\begin{lemma}\label{lem:isatree}
$T^-$ and $T^+$ are trees on at least three vertices with leaf sets $M^-$ and $M^+$, respectively. There is no vertex of degree two in $T^+$ and the same holds for $T^-$ if $V^- \neq \emptyset$.
\end{lemma}
\longversion{
\begin{proof}
First, assume that $V^- \neq \emptyset$ (the proof for $T^+$ is analogous). Then $H^-$ does not contain any chord of $C$. Consider any two inner faces $f$ and $g$ of $H^-$ that are incident to a common edge $e$. Since $e$ is no chord of $C$ and $C$ is isolating, $f$ and $g$ are incident to a exactly one common vertex $v \in V^-$. By construction of $T^-$, all vertices of $V^-$ that are incident to $f$ or $g$ are connected in $T^-$; in particular, $\deg_{T^-}(v) \geq \deg_{G}(v) \geq 3$. Hence, $T^-$ is connected. As $C$ is isolating, every two faces of $H^-$ are incident to at most one common vertex of $V^-$. Hence, the union of the acyclic graphs that are constructed for every major face of $H^-$, and thus $T^-$ itself, is acyclic. We conclude that $T^-$ is a tree with inner vertex set $V^-$, leaf set $M^-$ and no vertex of degree two.

Assume that $V^- = \emptyset$. Then $H^-$ is connected and outerplanar, and it is well-known that the weak dual of such a graph is a tree. By planar duality, $M^-$ is the set of vertices of degree one in $T^-$. It remains to show that $|V(T^-)| \geq 3$. By the previous result for $T^+$, $T^+$ contains at least three vertices and no vertex of degree two, so that $|M^+| \geq 3$. Consider any minor face $f \in M^+$ and its two extremal $C$-vertices $v$ and $w$. Since $G$ is polyhedral, $\{v,w\}$ is not a 2-separator of $G$, so that a non-extremal $C$-vertex of $f$ is incident to a chord of $C$ in $H^-$. Since $|M^+| \geq 3$, $H^-$ contains at least two chords of $C$, which gives $|V(T^-)| \geq 3$. In particular, no face of $H$ has boundary $E(C)$.
\end{proof}
}

\longversion{Note that $T^-$ may contain vertices of unbounded degree even if every vertex of $V^-$ has degree three in $G$ (for example, $\deg_{T^-}(h)=4$ in Figure~\ref{fig:IsolatingCycle2}).} We now relate the number of vertices in $V-V(C)$ to the number of minor faces of $H$.

\begin{lemma}\label{lem:treeminorfaces}
$|M^-| \geq |V^-| + 2$ and $|M^+| \geq |V^+| + 2$.
\end{lemma}
\begin{proof}
Consider the first claim $|M^-| \geq |V^-| + 2$. By Lemma~\ref{lem:isatree}, $|V(T^-)| \geq 2$. Every tree $T$ on at least two vertices has exactly $2 + \sum_{v \in V(T),\ \deg(v) \geq 3} \left( \deg(v)-2 \right)$ leaves. If $V^- = \emptyset$, this implies the claim directly. If $V^- \neq \emptyset$, the claim follows from the formula, as $T^-$ has no vertex of degree two by Lemma~\ref{lem:isatree}. The proof for $T^+ \neq \emptyset$ is analogous.
\end{proof}

By definition of minor faces, $H$ has no minor 0-face. Consider a minor 1-face $f$ of $H$ with $C$-edge $vw$; since $G$ is simple, $f$ is thick. Then the cycle obtained from $C$ by replacing $vw$ with the path $vv_fw$ shows that $C$ is extendable, which contradicts our assumption. We conclude that $H$ has no minor 1-face. This implies $c \geq 6$, since $V^+ \neq \emptyset$ implies $|M^+| \geq 3$ by Lemma~\ref{lem:treeminorfaces}. To summarize our assumptions so far, we know that $C$ is not extendable, $6 \leq c < \min\{\frac{2}{3}(n+3),n\}$, $|V^-| \leq |V^+| \geq 1$ and $H$ has no minor 1-face.

For the final contradiction to these assumptions, we aim to prove
\begin{align}
2c &\geq 4(|M^-| + |M^+|) \text{ if $V^- \neq \emptyset$ and}\label{eq:inequality1}\\
2c &\geq 2|M^-| + 4|M^+| \text{ if $V^- = \emptyset$.}\label{eq:inequality2}
\end{align}

This contradicts our assumption $c < \frac{2}{3}(n+3)$ by the following lemma.

\begin{lemma}\label{lem:inequalities}
Inequality~\eqref{eq:inequality1} implies $c \geq \frac{2}{3}(n+4)$ and Inequality~\eqref{eq:inequality2} implies $c \geq \frac{2}{3}(n+3)$.
\end{lemma}
\begin{proof}
By Lemma~\ref{lem:treeminorfaces}, $|M^-| \geq |V^-| + 2$ and $|M^+| \geq |V^+| + 2$. Moreover, we have $|V^-| + |V^+| = n-c$. Hence, if $V^- \neq \emptyset$, Inequality~\ref{eq:inequality1} implies $c \geq 2(n-c+4)$ and thus $c \geq \frac{2}{3}(n+4)$. If $V^- = \emptyset$, we have $|V^+| = n-c$, so that Inequality~\ref{eq:inequality2} implies $c \geq 2(n-c+3)$ and thus $c \geq \frac{2}{3}(n+3)$.
\end{proof}

We note that in the case $V^- \neq \emptyset$, Lemma~\ref{lem:inequalities} slightly strengthens the bound $\lfloor \frac{2}{3}(n+4) \rfloor$ of both the Isolation Lemma and Theorem~\ref{thm:essential} to $\frac{2}{3}(n+4)$.

In order to prove Inequality~\eqref{eq:inequality1} or~\eqref{eq:inequality2}, we will charge every $j$-face of $H$ with weight~$j$; hence, the total charge has weight $2c$. Then we discharge (i.e.\ move) these weights to minor faces such that no face has negative weight. We will prove that after the discharging every minor face of $H$ has sufficiently large weight (at least the coefficient given in the respective inequality) to satisfy Inequality~\eqref{eq:inequality1} or~\eqref{eq:inequality2}. The only problem are minor 2- and 3-faces, as these are charged with weight less than~4. We will transfer sufficient weight to them, so that the problem shifts to large minor faces, for which we then examine their (local and non-local) neighborhood in order to find that $C$ is extendable.

\subsection{Arches and Tunnels}
For a face $f$ of $H$, a path $A$ of $G$ is an \emph{arch} of $f$ if $f$ is minor and $A$ is either the maximal path in $H-E(C)$ all of whose edges are incident to $f$ (in this case we say that $A$ is \emph{proper}; then $A$ has length one or two depending on whether $f$ is thin or thick) or a chord of $C$ whose inner point set is strictly contained in $f$ and that does not join the two extremal $C$-vertices of $f$ (see Figure~\ref{fig:IsolatingCycle1}). \longversion{Hence, an arch $A$ is proper if and only if $A \subseteq H$, and every minor face $f$ has exactly one proper arch.}
Since $|V(T^-)| \geq 3$ by Lemma~\ref{lem:treeminorfaces}, no two leaves of $T^-$ are adjacent in $T^-$. Hence, every arch $A$ is an arch of exactly one face of $H$, which we call the \emph{face} $f(A)$ of $A$.

Let the \emph{archway} of an arch $A$ be the path in $C$ between the two end vertices of $A$ for which all edges are incident to $f(A)$. Since any arch $A$ and its archway bound a face $g$ in the graph $A \cup C$, we define $m_A := m_g$, $A$ as a $j$-\emph{arch} if $m_A = j$, $A$ as \emph{thick} if $f(A)$ is thick, and the (extremal) $C$-\emph{vertices} and $C$-\emph{edges} of $A$ as the (extremal) $C$-vertices and $C$-edges of $g$. \longversion{Note that every arch has exactly two extremal $C$-edges. By the last condition of the definition of arches, no two arches of $f$ have the same archway (in fact, the archways of the arches of a face form a laminar family on $E(C)$).}
If $f$ and $g$ are arches or faces of $H$, let $m_{f,g}$ be the number of $C$-edges that $f$ and $g$ have in common; we say that $f$ and $g$ are \emph{opposite} if $m_{f,g} > 0$.
An \emph{arch} $A$ \emph{of an arch} $B$ is an arch of $f(B)$ such that every $C$-edge of $A$ is a $C$-edge of $B$; we also say that $B$ \emph{has arch} $A$.

\begin{figure}[!htb]%
	\centering
	\captionbox{An acyclic counterclockwise tunnel track $T = (T_1,T_2,T_3,T_4,T_5)$ with exit pair $(g',e')$. Here, $(g',e')$ is on-track with itself, $(f(T_1),v_0v_1)$, $(h,v_2v_3)$, $(g,e)$, $(f,v_6v_7)$ and $(f(T_5),v_8v_9)$, but not with $(g,v_2v_3)$, which is on-track with $(f,v_8v_9)$. While $(f(T_1),v_0v_1)$, $(h,v_2v_3)$ and $(g,e)$ are transfer pairs, $(f,v_6v_7)$ and $(f(T_5),v_8v_9)$ are not: the former, as neither $v_6v_7$ nor $v_7v_8$ is an extremal $C$-edge of $f$; the latter, as $T_5$ has only one opposite face. \longversion{The transfer arches of $T$ are $T_1$, $T_2$ and $T_3$ (we color arches that are known to be transfer arches \emph{grey}).}
			\label{fig:On-Track}}[0.97\linewidth]
			{\includegraphics[page=1,scale=0.9]{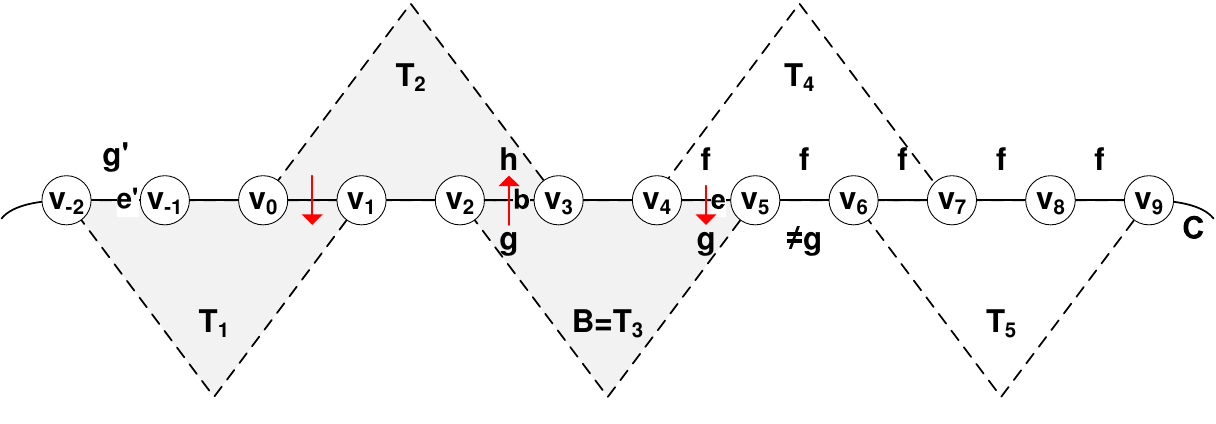}}
\end{figure}

Consider the 3-arches $T_1,\dots,T_k$ in Figure~\ref{fig:On-Track} and assume for now that every $T_i$ is thick and proper, so that every $f(T_i)$ is a minor 3-face. Since every $f(T_i)$ receives only initial weight~3 and $k$ is unbounded, every local method of transferring weights to reach weight at least~4 per minor face is bound to fail. Unfortunately, Figure~\ref{fig:On-Track} is not the only example where non-local methods are needed: in fact, there are infinitely many configurations in which weights must be transferred non-locally. We will therefore design the upcoming discharging rule in such a way that weight transfers do not depend on faces but instead on arches; this will reduce all structures that have to be handled non-locally to one common non-local structure (called tunnel), which is essentially the one shown in Figure~\ref{fig:On-Track}.

Let two 3-arches $A$ and $B$ be \emph{consecutive} if $m_{A,B}=1$. The reflexive and transitive closure of this symmetric relation partitions the set of all 3-arches whose middle $C$-edge is not a $C$-edge of a minor thin 2-face; we call the sets of this partition \emph{tunnels} (see Figure~\ref{fig:On-Track}). \longversion{Since $G$ is plane, $G$ imposes a notion of clockwise and counterclockwise on $C$; in the following, both directions always refer to $C$.} The \emph{counterclockwise track} of a tunnel $T$ (which will transfer weights counterclockwise around $C$) is the sequence $(T_1,T_2,\dots,T_k)$ of the 3-arches of $T$ such that $T_{i+1}$ is the clockwise consecutive successor of $T_i$ for every $1 \leq i < k$. We call a (counterclockwise or clockwise) tunnel track $(T_1,T_2,\dots,T_k)$ and its tunnel $T$ \emph{cyclic} if $k \geq 3$ and $T_k$ and $T_1$ are consecutive, and \emph{acyclic} otherwise.

The \emph{exit pair} $(g,e)$ of a counterclockwise track consists of the counterclockwise extremal $C$-edge $e$ of $T_1$ and the $e$-opposite face $g$ of $f(T_1)$. \emph{Clockwise tracks} and \emph{exit pairs} are defined analogously. The \emph{exit pairs} $(g,e)$ and $(g',e')$ of a tunnel $T$ are the two exit pairs of the counterclockwise and clockwise tracks of $T$; we call $g$ and $g'$ \emph{exit faces} of $T$. \longversion{We have $e = e'$ if and only if $T$ is cyclic, since $e = e'$ implies $k \neq 2$ due to $c \geq 6$; moreover, if $e = e'$, $g'$ and $g$ are opposite faces, so that $g \neq g'$. Hence, the exit pairs of an acyclic tunnel are always different, while its exit faces may be identical.}

In order to describe the weight transfers through tunnels, we define the following reflexive and symmetric relation for faces $g$ and $g'$ of $H$ and extremal $C$-edges $e$ and $e'$ of (not necessarily different) 3-arches of an acyclic tunnel $T$ such that $e$ and $e'$ are incident to $g$ and $g'$, respectively. Let $(g,e)$ be \emph{on-track with} $(g',e')$ if the following statements are equivalent (see Figure~\ref{fig:On-Track}).
\begin{itemize}
	\item $g$ and $g'$ are contained in the same region of $\mathbb{R}^2 - C$
	\item the distance between $e$ and $e'$ in the union of the $C$-edges of $T$ (measured by the length of a path that does not exceed $T$) is a multiple of $4$.
\end{itemize}

\longversion{
Clearly, this relation is an equivalence relation. Moreover, if $e$ is an extremal $C$-edge of a 3-arch $A$ of a tunnel $T$, $(f(A),e)$ is on-track with exactly one exit pair of $T$. Tunnels will serve as objects through which we can pull weight over long distances. We will later prove that tunnels transfer weights only one-way, i.e.\ towards the exit face of at most one of its tracks. Based on the structure of $G$, this weight may not be transferred through the whole tunnel track; the following definition restricts the parts where weight transfers may occur.
}

Let $T$ be an acyclic tunnel track, $e$ an extremal $C$-edge of an arch $B$ of $T$ such that $(g,e) := (f(B),e)$ is on-track with the exit pair of $T$, $b$ the extremal $C$-edge of $B$ different from $e$, and $h$ the $b$-opposite face of $g$ (see Figure~\ref{fig:On-Track}). Informally, $(h,b)$ is the on-track pair in $T$ that precedes $(g,e)$. Recursively, we define that $(g,e)$ is a \emph{transfer pair} of $T$ if $(h,b)$ is either the exit pair of $T$ or a transfer pair, and
\begin{itemize}
	\item $g$ is thick,
	\item the $e$-opposite face $f$ of $g$ is minor, $m_f \geq 3$, and $h \neq f$, and
	\item $e$ is either an extremal $C$-edge of $g$ or adjacent to such an edge, and in the latter case the middle $C$-edge of $B$ is either
	\begin{itemize}
		\item incident to $f$, or
		\item incident to a major face such that $e$ is an extremal $C$-edge of a 3-arch $A \neq B$ whose other extremal $C$-edge is not incident to a major face.
	\end{itemize}
\end{itemize}

\longversion{An arch $B$ of a tunnel track $T$ is called \emph{transfer arch} of $T$ if $(f(B),e)$ is a transfer pair of $T$, where $e$ is the extremal $C$-edge of $B$ such that $(f(B),e)$ is on-track with the exit pair of $T$. Note that a transfer arch of $T$ is not necessarily a transfer arch of the other track of the tunnel. A \emph{transfer arch of a tunnel} is an arch that is transfer arch of at least one of the tracks of the tunnel.}

\subsection{Discharging Rule}
By saying that a face $g$ \emph{pulls weight} $x$ \emph{over} its $C$-edge $e$ for a positive weight $x$, we mean that $x$ is added to $g$ and subtracted from the $e$-opposite face of $g$; we sometimes omit $x$ if the precise value is not important, but positive.

\begin{figure}[!htb]%
	\centering
	\subcaptionbox{Condition C2: $g$ is either a thick 2-face or a 3-face whose middle $C$-edge is incident to a thin minor 2-face $h \neq f$. The arrows depict that $g$ pulls weight~1 over $e$; we do not indicate weights pulled over other edges here. The vertex $v_g$ of a minor face $g$ is drawn only if (as here) $g$ is known to be thick. The \emph{red dotted} arches do not exist in $G$.
			\label{fig:C2}}[0.47\linewidth]
			{\includegraphics[page=1,scale=0.84]{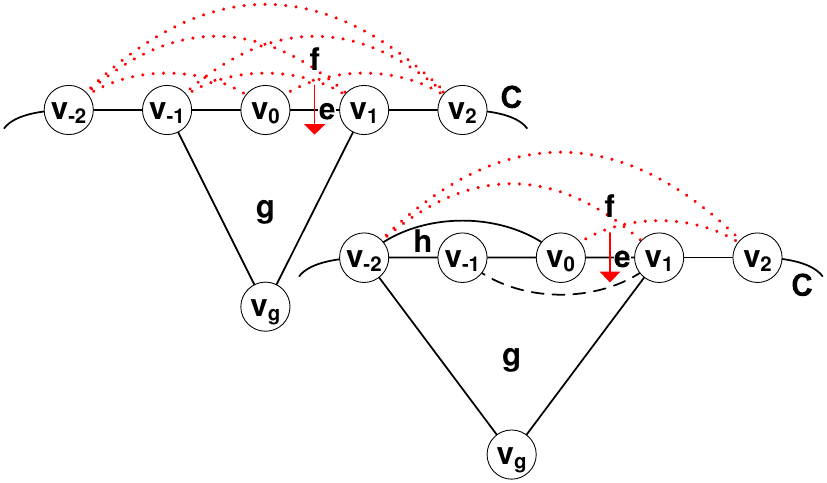}}
	\hspace{0.3cm}
	\subcaptionbox{Condition C3: $m_f \geq 3$, $B$ has no opposite major face, $e$ is not an extremal $C$-edge of a 3-arch of $f$, $v_1v_2$ is an extremal $C$-edge of $g$, and ($b$ is incident to a face $h \notin \{f,g\}$ or $m_g = 3$). Arches like $B$ that are not known to be proper (i.e.\ that are not known to be in $H$) are drawn \emph{dashed}. Note that $g$ may have more than three $C$-edges.
			\label{fig:C3}}[0.47\linewidth]
			{\includegraphics[page=1,scale=0.84]{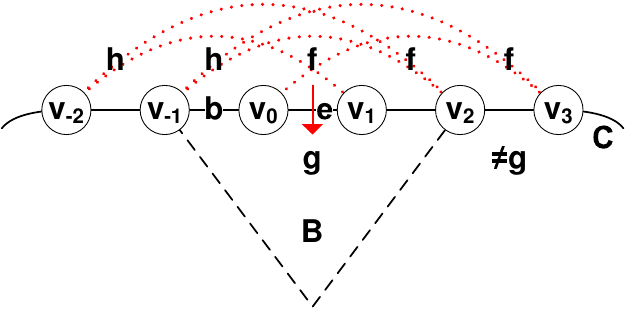}}
	\hspace{0.3cm}
	\subcaptionbox{Condition C4.
			\label{fig:C4}}[0.47\linewidth]
			{\includegraphics[page=1,scale=0.84]{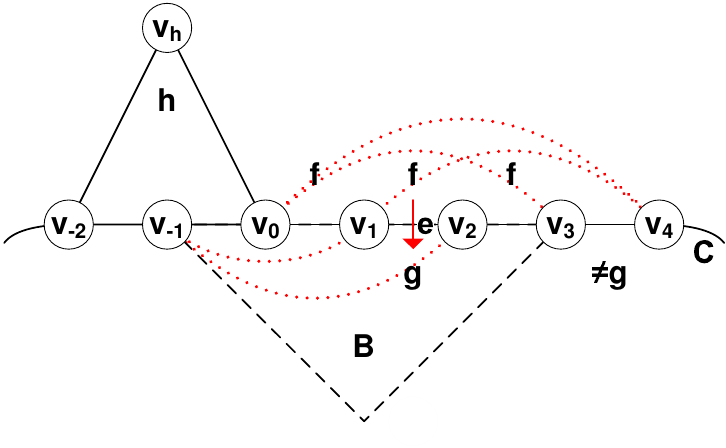}}
	\hspace{0.3cm}
	\subcaptionbox{Condition C5: $(h,b)$ is a transfer pair, no 2-arch of $g$ or $h$ has extremal $C$-vertex $v_{-1}$, and $e$ is not an extremal $C$-edge of a 3-arch of $f$ or of $g$.
			\label{fig:C5}}[0.47\linewidth]
			{\includegraphics[page=1,scale=0.84]{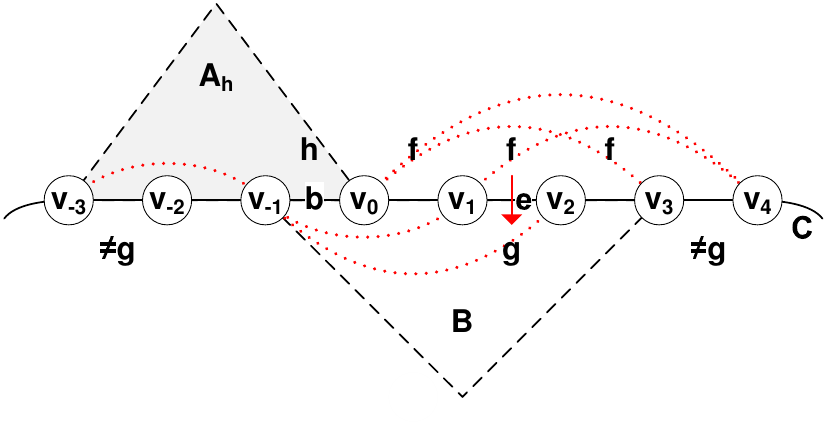}}
	\hspace{0.3cm}
	\subcaptionbox{Condition C6: $h \neq f$, and $e$ is not the middle $C$-edge of a 3-arch of $g$.
			\label{fig:C6}}[0.47\linewidth]
			{\includegraphics[page=1,scale=0.84]{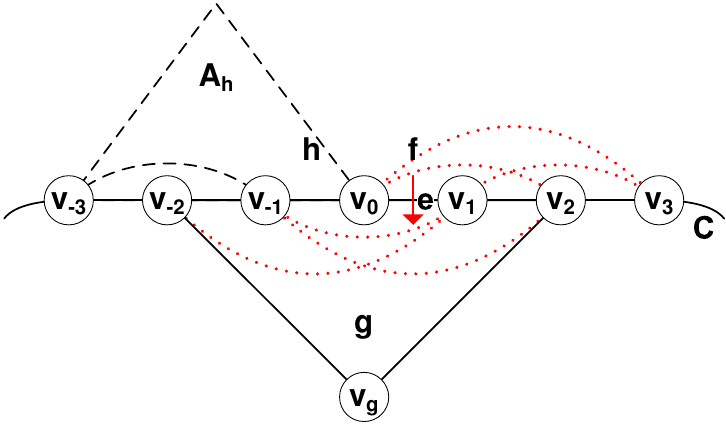}}
	\hspace{0.3cm}
	\subcaptionbox{Condition C7: $(g,e)$ is a transfer pair of the acyclic tunnel track $(T_1,T_2,T_3)$, and $(g',e')$ satisfies at least one of the conditions $C1$--$C6$.
			\label{fig:C7}}[0.47\linewidth]
			{\includegraphics[page=1,scale=0.84]{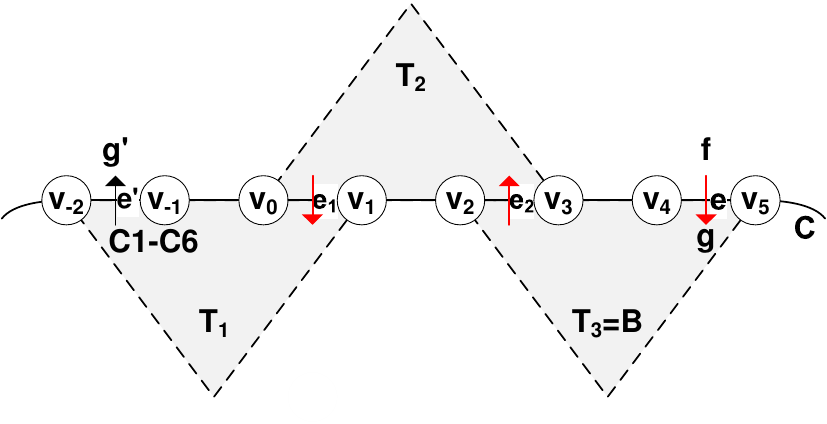}}
	\hspace{0.3cm}
	\caption{Conditions C2--C7}\label{fig:Conditions}
\end{figure}

\begin{definition}[Discharging Rule]\label{def:rules}
For every minor face $g$ of $H$ and every $C$-edge $e$ of $g$ (both in arbitrary order), $g$ pulls weight~1 over $e$ from the $e$-opposite face $f$ of $g$ for every of the following conditions that is satisfied (see Figure~\ref{fig:Conditions}).
\begin{enumerate}%
	\item[C1:] $f$ is major
	\item[C2:] $f$ is minor, and $g$ is either a thick 2-face or a 3-face whose middle $C$-edge is incident to a thin minor 2-face $h \neq f$
	\item[C3:] $f$ is minor and $m_f \geq 3$, $e$ is the middle $C$-edge of a 3-arch $B$ of $g$ and not an extremal $C$-edge of a 3-arch of $f$, $B$ has no opposite major face, an extremal $C$-edge of $B$ is an extremal $C$-edge of $g$, and (the other extremal $C$-edge of $B$ is incident to a face $h \notin \{f,g\}$ or $m_g = 3$)
	\item[C4:] $f$ is minor, $e$ is a non-extremal $C$-edge of a 4-arch $B$ of $g$ such that the extremal $C$-edge of $B$ that is adjacent to $e$ is an extremal $C$-edge of $g$, the other extremal $C$-edge of $B$ is incident to a thick minor 2-face $h$, and $m_{f,B} = 3$ %
	\item[C5:] $f$ is minor, $e$ is a non-extremal $C$-edge of a 4-arch $B$ of $g$ such that the extremal $C$-edge of $B$ that is adjacent to $e$ is an extremal $C$-edge of $g$, the other extremal $C$-edge $b$ of $B$ satisfies that $(h,b)$ is a transfer pair, the extremal $C$-vertex of $B$ that is incident to $b$ is not an extremal $C$-vertex of a 2-arch of $g$ or of $h$, $e$ is not an extremal $C$-edge of a 3-arch of $f$ or of $g$, and $m_{f,B} = 3$ %
	\item[C6:] $f$ is minor, $e$ is a non-extremal $C$-edge of a thick 4-face $g$ such that the extremal $C$-edge of $g$ that is not adjacent to $e$ is the middle $C$-edge of a 3-arch $A_h$ of a face $h \neq f$, and $e$ is not the middle $C$-edge of a 3-arch of $g$ %
	\item[C7:] $f$ is minor, $(g,e)$ is a transfer pair of an acyclic tunnel track $T$, and the exit pair $(g',e')$ of $T$ satisfies (in the notation $g$ and $e$) at least one of the conditions $C1$--$C6$
\end{enumerate}
\end{definition}

\longversion{Note that the weight transfers of this rule are solely dependent on $G$ and $C$ (and not on the current weight transfers). In particular, this holds for the ones caused by~$C7$, as these do not depend on other transfers caused by~$C7$, and the ones caused by $C5$, as these are only dependent on transfer pairs, not transfers. Since tunnels partition a subset of 3-arches, it suffices to evaluate~$C7$ once for each tunnel track after Conditions~$C1$--$C6$ have been evaluated.

After the discharging rule has been applied, $C7$ effectively routes weight~1 through a part of a tunnel track towards its exit face if this exit face pulls weight from $T$ by any other condition.
By definition of $C1$--$C6$, the only faces that do not have an arch of a tunnel $T$ (i.e.\ reside ``outside'' $T$) but pull weight over a $C$-edge of such an arch are the exit faces of $T$; in this sense, weight may leave $T$ only through an exit face of $T$.
}

\subsection{Structure of Tunnels and Transfers}
We give further insights into the structure of tunnels and the location of edges over which our discharging rule pulls (positive) weight.

\begin{lemma}\label{lem:NoAcyclicTunnelOverlap}
Every tunnel track $(T_1,\dots,T_k)$ with $k \geq 3$ satisfies $m_{T_1,T_k}=0$. In particular, every tunnel is acyclic.
\end{lemma}
\longversion{\begin{proof}
Assume first that $(T_1,\dots,T_k)$ is cyclic, i.e.\ that $m_{T_1,T_k}=1$. Since $k \geq 3$, this implies $c=2k$. Consider the middle $C$-edge $e$ of any $T_i$ and let $g$ be the $e$-opposite face of $f(T_i)$. If $g$ is minor, $g$ is the face of some $T_j \neq T_i$ of $T$, as otherwise $g$ would be a minor 1-face of $H$, which contradicts our assumptions. Hence, $H$ has at most $k$ minor faces, so that Inequality~\eqref{eq:inequality1} holds. This implies $c \geq \frac{2}{3}(n+4)$ by Lemma~\ref{lem:inequalities}, which contradicts our assumptions.

Hence, $m_{T_1,T_k} \neq 1$. No two 3-arches have the same set of extremal $C$-vertices, as such a set would be a 2-separator of $G$, which contradicts that $G$ is polyhedral. Thus, $m_{T_1,T_k} \in \{0,2\}$. Assume to the contrary that $m_{T_1,T_k} = 2$. Since $k \geq 3$, this implies $c=2k-1$. As $G$ is polyhedral, the $C$-vertex $v_1$ of $T_1$ (in the notation of Figure~\ref{fig:NoAcyclicTunnelOverlap}) has degree at least three in $G$. Thus, $v_{-1}v_1 \in E(G)$ or $v_1v_3 \in E(G)$, say the latter by symmetry. This implies that the face $f$ of $T_1$ is thick, so that the vertex $v_f$ exists. If $f$ is a 3-face (i.e.\ $T_1$ is proper), $C$ is extendable, as the cycle obtained from $C$ by replacing the path $v_{-1}v_0v_1v_2v_3$ with the path $v_{-1}T_kv_2v_1v_0v_fv_3$ shows (this adds one or two new vertices to $C$).

\begin{figure}[!htb]
	\centering
	\captionbox{$m_f=4$: The fat path replacement shows that $C$ is extendable.
			\label{fig:NoAcyclicTunnelOverlap}}[0.97\linewidth]
			{\includegraphics[page=1,scale=0.9]{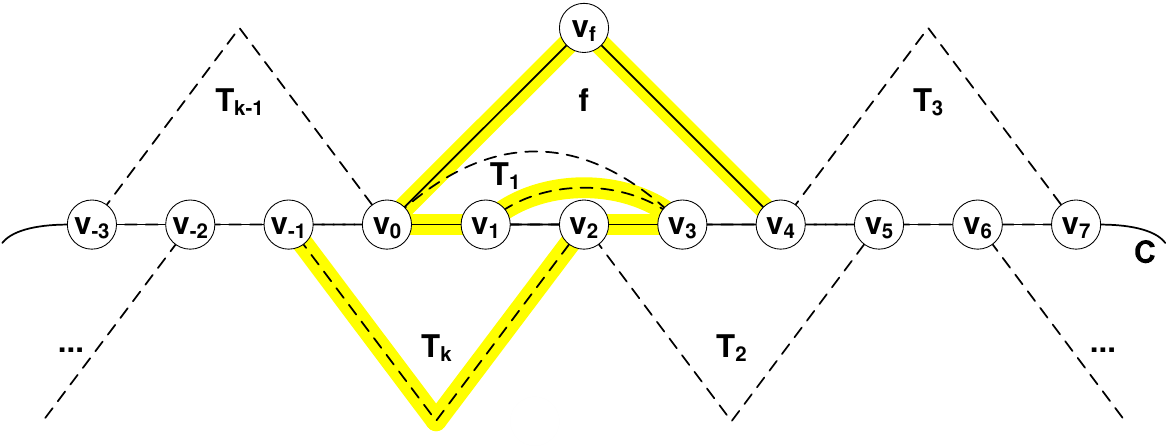}}
\end{figure}

Since this contradicts our assumption, $m_f \geq 4$. If $m_f = 4$, $v_3v_4$ is a $C$-edge of $f$, since $G$ is plane. Then $C$ is extendable by the path replacement shown in Figure~\ref{fig:NoAcyclicTunnelOverlap}. In the remaining case $m_f \geq 5$, planarity implies that $T_3$ or $T_{k-1}$ is a 3-arch of $f$. Then $H$ has at most $k-1$ minor faces, so that Inequality~\eqref{eq:inequality1} holds, which contradicts our assumptions by Lemma~\ref{lem:inequalities}.
\end{proof}}

\longversion{We remark that it is possible, but more involved, to prove Lemma~\ref{lem:NoAcyclicTunnelOverlap} solely by using the discharging rule of Definition~\ref{def:rules}.} Using Lemma~\ref{lem:NoAcyclicTunnelOverlap}, we assume from now on that every tunnel is acyclic. We next show that $G$ does not contain the dotted arches of Figure~\ref{fig:Conditions} for the respective conditions; this sheds first light on the implications that are triggered by the assumption that $C$ is not extendable.

\begin{lemma}\label{lem:conditionrestrictions}
For any satisfied condition~$X \in \{C2,\dots,C6\}$, none of the red dotted arches in the respective Figure~\ref{fig:C2}--\ref{fig:C6} exist in the depicted face of $H$. If $X = C2$ and $g$ is a 3-face, $v_{-1}v_1 \in E(G)$; if $X = C6$, $v_{-3}v_{-1} \in E(G)$.
\end{lemma}
\longversion{
\begin{proof}
We use the notation of Figure~\ref{fig:Conditions}.
Assume $X = C2$. First, let $g$ be a 2-face in Figure~\ref{fig:C2}. If $v_0v_1$ (or, by symmetry, $v_{-1}v_0$) is a $C$-edge of a 2-arch $A$ of $f$, $v_0$ is an extremal $C$-vertex of $A$, since $\{v_{-1},v_1\}$ is not a 2-separator of $G$; then $C$ is extendable by the path replacement $v_{-1}v_gv_1v_0Av_2$ (this adds one or two new vertices to $C$, depending on whether $A$ is proper and thick), which contradicts our assumptions. If $v_0v_1$ (or, by symmetry, $v_{-1}v_0$) is the middle $C$-edge of a 3-arch, we have $v_0v_2 \in E(G)$, as $G$ is polyhedral and thus $\deg_G(v_0) \geq 3$; since this contradicts the previous result, neither $v_0v_1$ nor $v_{-1}v_0$ is the middle $C$-edge of a 3-arch. Using the same argument, $v_0$ is not the middle $C$-vertex of any 4-arch.

Let $g$ be a 3-face. Since $G$ is polyhedral, we have $\deg_G(v_{-1}) \geq 3$ and thus $v_{-1}v_1 \in E(G)$ by planarity. If $v_0v_2 \in E(G)$, $C$ is extendable by the path replacement $v_{-2}v_gv_1v_{-1}v_0Av_2$ (adding exactly one new vertex to $C$). Since $h$ is thin, $f$ is thin, so that $v_0v_2 \notin E(G)$ implies that no arch has extremal $C$-vertices $v_0$ and $v_2$. Since neither $\{v_{-2},v_1\}$ nor $\{v_{-2},v_2\}$ is a 2-separator of $G$ and $v_0v_2 \notin E(G)$, no face different from $g$ has an arch whose set of extremal $C$-vertices is $\{v_{-2},v_1\}$ or $\{v_{-2},v_2\}$.

Assume $X = C3$. By definition of~$C3$, $v_0v_1$ is not an extremal $C$-edge of a 3-arch of $f$. Since $G$ is polyhedral, $v_0v_1$ is not the middle $C$-edge of a 3-arch of $f$. Assume to the contrary that $v_1$ (or $v_0$ by a symmetric argument) is the middle $C$-vertex of a 4-arch of $f$. Then $v_{-1}v_0$ is incident to $f$, which implies by~$C3$ that $g$ is a minor 3-face. Because $\{v_{-1},v_2\}$ is not a 2-separator of $G$, then a vertex of $\{v_0,v_1\}$ is adjacent to $v_3$ in $G$. Since $v_0v_3 \notin E(G)$, we have $v_1v_3 \in E(G)$ and thus $v_0v_2 \in E(G)$ by $\deg_G(v_0) \geq 3$. Then $g$ is thick and $C$ is extendable by Figure~\ref{fig:RestrictionC3}.

\begin{figure}[!htb]
	\centering
	\subcaptionbox{$C3$ when $v_1$ is the middle $C$-vertex of a 4-arch of $f$.
			\label{fig:RestrictionC3}}[0.47\linewidth]
			{\includegraphics[page=1,scale=0.9]{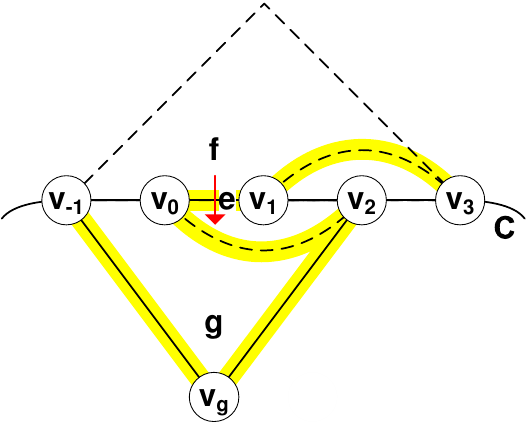}}
	\hspace{0.3cm}
	\subcaptionbox{$C4$ when $v_2$ is the middle $C$-vertex of a 4-arch of $f$.
			\label{fig:RestrictionC4}}[0.47\linewidth]
			{\includegraphics[page=1,scale=0.9]{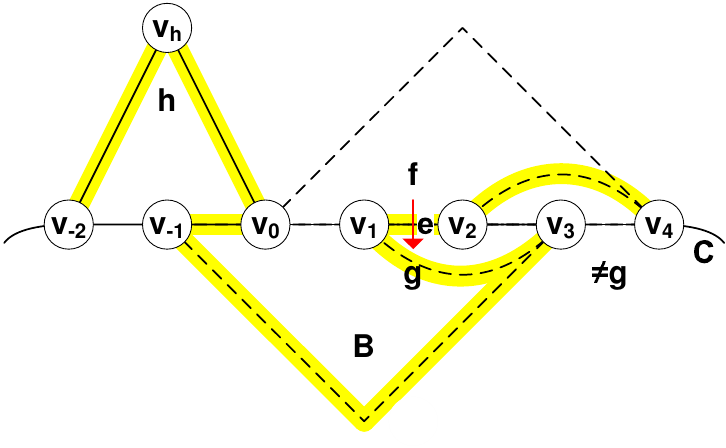}}
	\hspace{0.3cm}
	\caption{Restrictions of~$C3$ and~$C4$
	}\label{fig:restrictions}
\end{figure}

Assume $X = C4$. Then the first result of the case $X=C2$ implies $v_{-1}v_1 \notin E(G)$. In addition, $v_{-1}v_2 \notin E(G)$, as otherwise $C$ is extendable by the path replacement $v_{-2}v_hv_0v_1v_2v_{-1}Bv_3$. Hence, $v_1v_2$ is not an extremal $C$-edge of a 3-arch of $g$. If $v_1v_2$ is an extremal $C$-edge of a 3-arch $A$ of $f$, $C$ is extendable by the path replacement $v_{-2}v_hv_0v_{-1}Bv_3v_2v_1Av_4$, as this adds at most three new vertices to $C$.
Note that, if $A$ is proper, we have $v_A \neq v_h$ in this replacement ($A$ is thick, because $h$ is), as $H$ has no minor 1-face. %
In addition, $v_1v_2$ is not the middle $C$-edge of a 3-arch of $f$, as otherwise $\{v_0,v_3\}$ would be a 2-separator of $G$ by the previous results.
If $v_2$ is the middle $C$-vertex of a 4-arch of $f$, we have $v_2v_4 \in E(G)$, as otherwise $\{v_0,v_3\}$ would be a 2-separator of $G$. Since $\deg_G(v_1) \geq 3$, this implies $v_1v_3 \in E(G)$, so that $C$ is extendable by Figure~\ref{fig:RestrictionC4}.

Assume $X = C5$. By definition of transfer arches, $4 \leq m_g \leq 5$. If $v_1v_2$ is the middle $C$-edge of a 3-arch of $f$, $\{v_0,v_3\}$ is a 2-separator of $G$, since $v_{-1}v_1$ and $v_{-1}v_2$ are not contained in $G$. This contradicts that $G$ is polyhedral. Assume to the contrary that $v_2$ is the middle $C$-vertex of a 4-arch $A$ of $f$. If $m_g = 4$, $C$ is extendable by the replacement $v_{-1}v_gv_3v_2v_1v_0Av_4$, so let $m_g = 5$. Then $v_2v_4 \notin E(G)$, as otherwise $C$ is extendable by the replacement $v_{-2}v_gv_3v_{-1}v_0v_1v_2v_4$. Then $\{v_0,v_3\}$ is a 2-separator of $G$, which contradicts that $G$ is polyhedral.

Assume $X = C6$. Then $v_{-1}v_1 \notin E(G)$, as otherwise $C$ is extendable by the replacement $v_{-3}A_hv_0v_1v_{-1}v_{-2}v_gv_2$, as $v_g$ exists since $g$ is thick. Since $G$ is polyhedral, $\deg_G(v_{-1}) \geq 3$, which implies $v_{-3}v_{-1} \in E(G)$ as only remaining option. Then $v_{-2}v_1 \notin E(G)$, as otherwise $C$ is extendable by $v_{-3}v_{-1}v_0v_1v_{-2}v_gv_2$. Since $G$ is polyhedral, this implies that there is no 2-arch of $f$ with middle $C$-vertex $v_1$. Assume to the contrary that $f$ has a 2-arch $A$ with middle $C$-vertex $v_2$. Then $A$ is thick, as $A_h$ has a 2-arch, and $A$ is not proper, as otherwise $f$ would be a minor 1-face of $H$ due to $h \neq f$. Hence, $v_1v_3 \in E(G)$, so that $C$ is extendable by the replacement $v_{-3}A_hv_0v_{-1}v_{-2}v_gv_2v_1v_3$. In addition, $v_1v_2$ is not the middle $C$-edge of a 3-arch, as otherwise $v_1$ would have degree two in $G$.
\end{proof}
}

For a $C$-edge $e$ of a face $g$ of $H$ and a condition $X \in \{C1,C2,\dots,C7\}$, let \pulls{g}{e}{X} denote that~$X$ is satisfied for $g$ and $e$ in Definition~\ref{def:rules}.
\longversion{For notational convenience throughout this paper, whenever $g$ pulls weight from a face $f$, we denote by $v_0$ the extremal $C$-vertex of $f$ whose clockwise neighbor $v_1$ in $C$ is $C$-vertex of $f$, and denote by $v_i$ the $i$th vertex modulo $c$ in a clockwise traversal of $C$ starting at $v_1$\longversion{ (see for example Figure~\ref{fig:NoAcyclicTunnelOverlap})}.

So far, a tunnel might transfer weights through both of its tracks simultaneously. The next lemma shows that this never happens.

\begin{lemma}\label{lem:AtMostOnePullFromTunnel}
Let $(g,e)$ and $(g',e')$ be the exit pairs of a tunnel $T$ such that $g$ pulls weight over $e$. Then the following statements hold.
\begin{enumerate}
	\item $g$ is minor, \pulls{g}{e}{C2} and no other condition is satisfied for $(g,e)$\label{enum:tunnelC2}
	\item every 2-arch $A$ of an arch of $T$ has a $C$-edge $b$ such that $(f(A),b)$ is on-track with $(g,e)$\label{enum:2archOnTrack}
	\item $g \neq g'$ and there is no 2-arch of $g'$ that has $C$-edge $e'$\label{enum:no2arch}
	\item $g'$ does not pull any weight over $e'$\label{enum:oneway}
	\item for every 4-arch $A$ that has an arch $T_i$ of $T$, the common extremal $C$-edge $b$ of $A$ and $T_i$ satisfies that $(f(A),b)$ is on-track with $(g,e)$\label{enum:no4arch}
	\item every arch $T_i$ of $T$ that is consecutive to two transfer arches of $T$ satisfies $m_{f(T_i)} \leq 4$\label{enum:3or4arch}
\end{enumerate}
\end{lemma}
\begin{proof}
Let $X$ be a condition in $\{C1,C2,\dots,C7\}$ such that \pulls{g}{e}{X}. Without loss of generality, assume that $(g,e)$ is the counterclockwise exit pair of $T$. Let $(T_1,\dots,T_k)$ be the counterclockwise track of $T$. For every $1 \leq j \leq k$, let $e_j$ denote the edge that joins the two extremal $C$-vertices of $T_j$.

For Claim~\ref{enum:tunnelC2}, $g$ is minor, because major faces do not pull weight over any edge. Since $f(T_1)$ is minor by the definition of arches, $X \neq C1$. If $X = C3$, $e$ is an extremal $C$-edge of the 3-arch $T_1$, which contradicts the definition of $C3$. If $X \in \{C4,C5,C6\}$, $T_1$ contradicts Lemma~\ref{lem:conditionrestrictions}. Assume $X=C7$. Then $e$ is an extremal $C$-edge of a transfer arch $A$ of $g$, which implies that $A$ and $T_1$ are consecutive. Hence, $A$ is an arch of the same tunnel $T$ as $T_1$, so that $A = T_k$ and $k \geq 3$ (the latter due to $c \geq 6$). Hence, $T$ is cyclic, which contradicts Lemma~\ref{lem:NoAcyclicTunnelOverlap} (and the definition of~$C7$). We conclude that $X = C2$. Note that in the case $m_g = 3$ of $C2$, the proper arch of $g$ is by definition not part of any tunnel.

\begin{figure}[!htbp]
	\centering
	\subcaptionbox{A 2-arch $A$ of $T_i$ that has no $C$-edge $b$ such that $(f(A),b)$ is on-track with $(g,e)$. Here, $G$ contains the edge $e_j$ joining the two extremal $C$-vertices of $T_j$ for every $j < i$.
			\label{fig:No2Arches1}}[0.97\linewidth]
			{\includegraphics[page=1,scale=0.9]{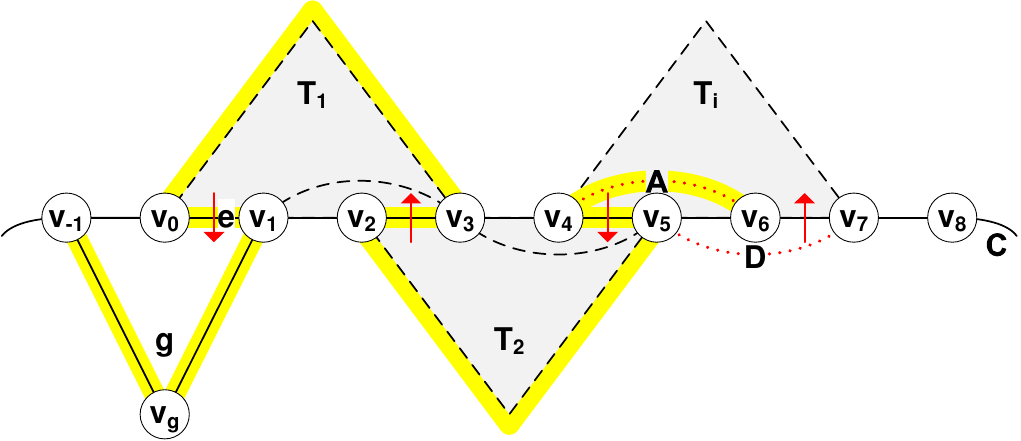}}
	\hspace{0.3cm}
	\subcaptionbox{$T_l = T_1$ has a 2-arch $B$, and $T_j = T_2$ has maximal $j < i$ such that $G$ does not contain $e_j$.
			\label{fig:No2Arches2}}[0.97\linewidth]
			{\includegraphics[page=1,scale=0.9]{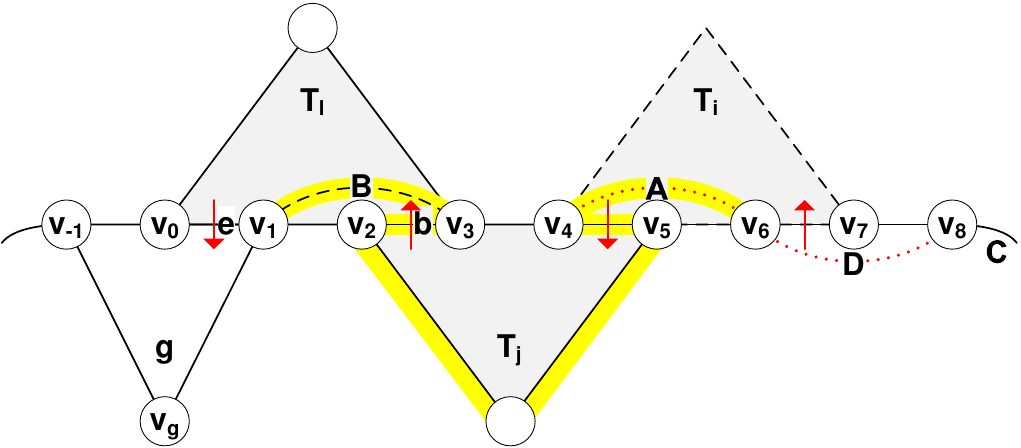}}
	\hspace{0.3cm}
	\subcaptionbox{$g=g'$. Here, the fat subgraph depicts the whole cycle $C'$ that replaces $C$. Note that this case occurs only one step before the desired cycle length is reached.%
			\label{fig:TunnelC2SameExitFace}}[0.97\linewidth]
			{\includegraphics[page=1,scale=0.9]{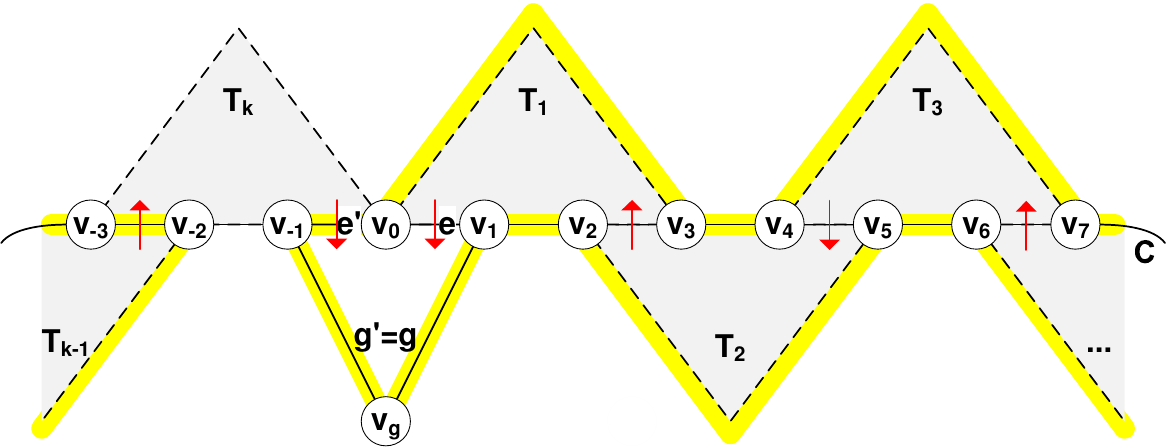}}
	\hspace{0.3cm}
	\subcaptionbox{A 4-arch $A$ having an arch $T_i \in T$ such that $(f(A),b)$ is not on-track with $(g,e)$.
			\label{fig:No4Arches}}[0.97\linewidth]
			{\includegraphics[page=1,scale=0.9]{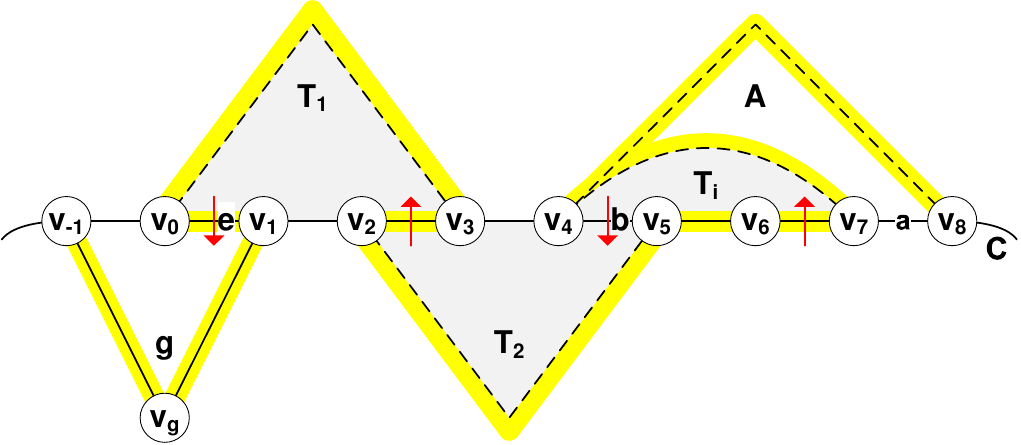}}
	\hspace{0.3cm}
	\caption{Exit face $g$ pulling weight from a tunnel such that $m_g = 2$. For the case $m_g = 3$, $v_{-1}v_gv_1v_0$ is replaced with $v_{-2}v_gv_1v_{-1}v_0$. %
	}\label{fig:AtMostOnePullFromTunnel}
\end{figure}

For Claim~\ref{enum:2archOnTrack}, assume to the contrary that some $T_i \in T$ has a 2-arch $A$ that has no $C$-edge $b$ such that $(f(A),b)$ is on-track with $(g,e)$; without loss of generality, let $i$ be minimal (see Figure~\ref{fig:No2Arches1}). By Claim~\ref{enum:tunnelC2}, $X = C2$. By Lemma~\ref{lem:conditionrestrictions} (for~$C2$), $i \neq 1$.

Assume first that, for every $j < i$, $G$ contains the edge $e_j$. Note that this happens precisely if either $e_j = T_j$ (i.e.\ $T_j$ is thin or non-proper) or $\{e_j\} \cup T_j$ bounds a triangle whose interior point set is contained in $f(T_j)$ (as $G$ is polyhedral). If $A$ and $g$ do not have any $C$-vertex in common, $C$ is extendable by the path replacement shown in Figure~\ref{fig:No2Arches1}, as this adds exactly one new vertex to $C$ (namely, $v_g$). Otherwise, $m_g = 2$ by planarity, and $A$ and $g$ have exactly $v_{-1}$ in Figure~\ref{fig:No2Arches1} as common $C$-vertex, which implies $i = k$ and $g = g'$. Then the same replacement (this time specifying the whole cycle that replaces $C$) shows that $C$ is extendable.

In the remaining case, $T$ has an arch $T_j$ such that $j < i$ is maximal and $G$ does not contain $e_j$; in particular, $T_j$ is thick and proper. Define $T_0 := g$ and let $0 \leq l \leq j$ be maximal such that $T_l$ has a 2-arch $B$ (see Figure~\ref{fig:No2Arches2}). Such $l$ exists, as $l=0$ is a valid choice for the case $m_g=2$ and, by Lemma~\ref{lem:conditionrestrictions}, also for the case $m_g=3$. By minimality of $i$, $B$ has a $C$-edge $b$ such that $(f(B),b)$ is on-track with $(g,e)$ (if $l=0$, we have $(f(B),b) = (g,e)$).

Consider the path replacement of Figure~\ref{fig:No2Arches2}, which contains besides edges of $C$ only $A$, $B$, arches $T_z$ and edges $e_z$ that satisfy $l < z < i$. For every $j < z < i$, the maximality of $j$ implies that $G$ contains $e_z$, so that taking $e_z$ in the replacement does not add any new vertex to $C$.

For every $i < z < j$, we take the edge $e_z$ if $G$ contains it; if so, this does not add a new vertex to $C$. If otherwise $e_z \notin E(G)$, $T_z$ is thick and proper and has no 2-arch by maximality of $l$, so that $f(T_z)$ is a face of size five in $G$. If $l < j$, $T_j$ has no 2-arch, so that $f(T_j)$ is a face of size five in $G$ for the same reason. We conclude that, if $l < j$, the path replacement adds at most $1+n_5(G)$ ($1$ because of $g$) new vertices to $C$, so that $C$ is extendable. If otherwise $l = j$, $C$ is extendable by restricting the replacement to $v_2T_jv_5Bv_3v_4e_{j+1}\dots e_{i-1}v_5v_4A$, which adds again at most $1+n_5(G)$ ($1$ because of $T_j$) new vertices to $C$. Note that every vertex (in particular, the ones not in $C$) is visited at most once by the replacement path, as otherwise $H$ would have a minor 1-face.

For the first statement of Claim~\ref{enum:no2arch}, assume to the contrary that $g=g'$ (see Figure~\ref{fig:TunnelC2SameExitFace}). By Claim~\ref{enum:tunnelC2}, $X=C2$. Since $g = g'$ pulls weight over both edges $e$ and $e'$ due to~$C2$, we may apply Claim~\ref{enum:2archOnTrack} for both exit pairs of $T$. This implies that no arch of $T$ has a 2-arch. Consider the fat cycle of Figure~\ref{fig:TunnelC2SameExitFace} that replaces $C$ by omitting an arbitrary arch of $T$ (in Figure~\ref{fig:TunnelC2SameExitFace}, $T_k$ is omitted). Then for every arch $T_z$ that is not omitted in this replacement, either $G$ contains $e_z$ or $f(T_z)$ is a face of size five in $G$. Hence, the replacement adds at most $1+n_5(G)$ ($1$ because of $g$) new vertices to $C$, so that $C$ is extendable.

For the second statement of Claim~\ref{enum:no2arch}, assume to the contrary that there is a 2-arch $D$ of $g'$ that has $C$-edge $e'$. If $m_{D,T_k} = 2$ (see Figure~\ref{fig:No2Arches1} when $i = k$), there is a 2-arch $A$ of $T_k$ that contradicts Claim~\ref{enum:2archOnTrack}, as $G$ has no vertex of degree two. Hence, $m_{D,T_k} = 1$ (see $D$ in Figure~\ref{fig:No2Arches2} when $i = k$). If $D$ has a common $C$-edge with $g$, we have $m_g = 2$ by planarity and $m_{D,g} \neq 1$ by Lemma~\ref{lem:conditionrestrictions}; this implies $g = g'$, which contradicts the previous claim. In the remaining case, $D$ and $g$ have no common $C$-edge. Then $C$ is extendable by the same path replacements of Figures~\ref{fig:No2Arches1} and~\ref{fig:No2Arches2} as in Claim~\ref{enum:2archOnTrack}, except that this adds at most $2+n_5(G)$ new vertices to $C$ (i.e.\ at most one more), because $D$ may be thick and proper.

For Claim~\ref{enum:oneway}, assume to the contrary that there is a condition $Y \in \{C1,\dots,C7\}$ such that \pulls{g'}{e'}{Y}. By Claim~\ref{enum:tunnelC2}, $X = Y = C2$. This contradicts Claim~\ref{enum:no2arch} in both cases $m_g = 2$ and $m_g = 3$.

For Claim~\ref{enum:no4arch}, assume to the contrary that a 4-arch $A$ has some $T_i \in T$ such that the common extremal $C$-edge $b$ of $A$ and $T_i$ does not satisfy that $(f(A),b)$ is on-track with $(g,e)$ (see Figure~\ref{fig:No4Arches}). Let $a$ be the $C$-edge of $A$ that is not a $C$-edge of $T_i$. Assume first that $a$ is a $C$-edge of $g$. Then $m_g = 2$, $i = k$, $c = 2k+2$ and $e'$ is not incident to a minor 2-face by Claim~\ref{enum:no2arch}. Since $H$ has no minor 1-face, $H$ has thus at most $k+1$ minor faces, so that Inequality~\eqref{eq:inequality1} holds. This implies $c \geq \frac{2}{3}(n+4)$ by Lemma~\ref{lem:inequalities}, which contradicts our assumptions. Hence, $a$ is not a $C$-edge of $g$.

Consider the replacement of Figure~\ref{fig:No4Arches} and note that this replacement is also valid when $A$ and $g$ share exactly one $C$-vertex. If $G$ contains $e_j$ for every $j < i$, $C$ is extendable by this replacement, as this adds at most $2$ new vertices to $C$. Otherwise, we may proceed as in the proof of Claim~\ref{enum:2archOnTrack} and compensate the usage of new vertices that are added to $C$ with faces of size five of $G$, so that at most $2+n_5(G)$ ($2$ because of $A$ and either $T_j$ or $g$) new vertices are used.

\begin{figure}[!htbp]
	\centering
	\captionbox{A 5-arch $A$ having exactly the $C$-edges of $T_i$ as non-extremal $C$-edges.
			\label{fig:3or4arch}}[0.97\linewidth]
			{\includegraphics[page=1,scale=0.9]{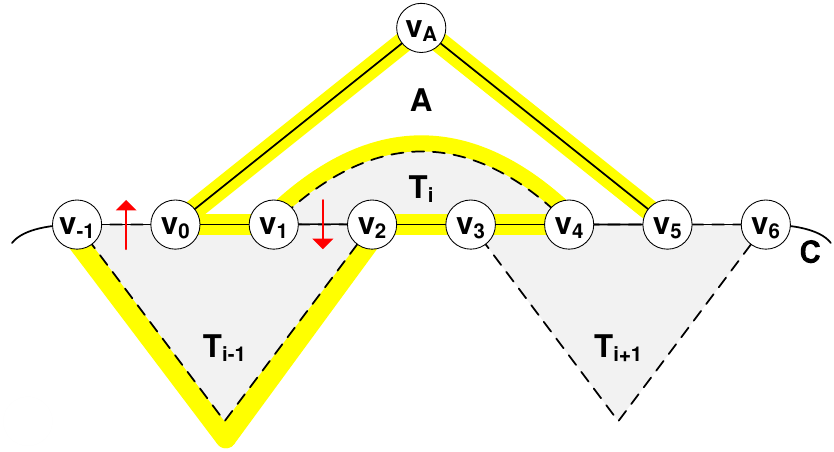}}
	\hspace{0.3cm}
\end{figure}

For Claim~\ref{enum:3or4arch}, assume to the contrary that $T$ contains an arch $T_i \notin \{T_1,T_k\}$ with minimal $i$ and $m_{f(T_i)} \geq 5$ such that $T_{i-1}$ and $T_{i+1}$ are transfer arches. If every $C$-edge of $T_{i+1}$ is incident to $f(T_i)$, $T_{i+1}$ contradicts the definition of transfer arches, as all its $C$-edges are opposite to the same face. By the same argument, not every $C$-edge of $T_{i-1}$ is incident to $f(T_i)$. We conclude that $m_{f(T_i)} = 5$ and that the $C$-edges of $T_i$ are exactly the non-extremal $C$-edges of $f(T_i)$. Then $C$ is extendable by Figure~\ref{fig:3or4arch}.
\end{proof}}

\longversion{By Lemma~\ref{lem:AtMostOnePullFromTunnel}\ref{enum:oneway},}\shortversion{One can prove that} all weight transfers that are caused within a tunnel by Condition~$C7$ are one-way, i.e.\ use only one track of $T$.\longversion{As an immediate implication, the following lemma shows that all weight transfers strictly within a tunnel are solely dependent on the weight transfers on its exit pairs.

\begin{lemma}\label{lem:ExitPairsDominateTunnelInside}
Let $(g,e)$ be a transfer pair of a tunnel track $T$ with exit pair $(g',e')$. Then $g$ pulls weight over $e$ if and only if $g'$ pulls weight over $e'$ (and if so, \pulls{g}{e}{C7} and \pulls{g'}{e'}{C2} such that $m_{g'} = 2$).
\end{lemma}
\begin{proof}
Assume that $g'$ pulls weight over $e'$. By Lemma~\ref{lem:NoAcyclicTunnelOverlap}, $T$ is acyclic. By Lemma~\ref{lem:AtMostOnePullFromTunnel}\ref{enum:tunnelC2}, \pulls{g'}{e'}{C2}; in particular, $2 \leq m_{g'} \leq 3$. Hence, $C7$ is satisfied for $(g,e)$, so that $g$ pulls weight over $e$. Assume to the contrary that $m_{g'} = 3$ and let $B$ be the first 3-arch of $T$. Then $B$ is thin and thus no transfer arch of $T$, which contradicts that $(g,e)$ is a transfer pair of $T$. Hence, $m_{g'} = 2$.

Assume to the contrary that \pulls{g}{e}{X} for some $X \in \{C1,\dots,C7\}$ and $g'$ does not pull any weight over $e'$. The latter implies $X \neq C7$. Since $(g,e)$ is a transfer pair, the $e$-opposite face of $g$ is minor. Hence, $X \neq C1$. Since $e$ is a $C$-edge of a 3-arch $B$ of $T$ with $f(B)=g$, $m_g \geq 3$; as $B$ is contained in a tunnel, $g$ is not a minor 3-face whose middle $C$-edge is incident to a thin minor 2-face. Hence, $X \neq C2$. By planarity, $X \neq C3$. Since $e$ is an extremal $C$-edge of $B$, Lemma~\ref{lem:conditionrestrictions} implies $X \notin \{C4,C5,C6\}$, which is a contradiction.
\end{proof}}
An immediate implication of the discharging rule in Definition~\ref{def:rules} is that every face pulls a non-negative integer weight over every edge, as every satisfied condition adds~1 to that weight. We next prove that no two of the conditions~$C1$--$C7$ are satisfied simultaneously for the same face $g$ and edge $e$; hence, $g$ pulls either weight~0 or~1 over $e$. \longversion{This is crucial for keeping the amount of upcoming arguments on a maintainable level; in fact, our conditions were designed that way.}

\begin{lemma}\label{lem:atMost1OverAnyEdge}
The total weight pulled by a face of $H$ over its $C$-edge $e$ is either~0 or~1. If it is~1, the $e$-opposite face does not pull any weight over $e$.
\end{lemma}
\longversion{
\begin{proof}
Assume to the contrary that $e$ is incident to two faces $f$ and $g$ of $H$ such that \pulls{f}{e}{X} and \pulls{l}{e}{Y} for conditions~$X$ and $Y$ and $l \in \{f,g\}$; without loss of generality, we assume that $Y$ is not stated before $X$ in Definition~\ref{def:rules}. In general, $X = Y$ implies $l = g$. If $X = C1$, $g$ is major, which implies $Y = C1$ and thus $l = g$; then $f$ is major, which contradicts \pulls{f}{e}{X}. Hence, $X \neq C1$, so that both $f$ and $g$ are minor.

If $Y = C7$, $e$ is an extremal $C$-edge of a 3-arch. Then $X \notin \{C3,\dots,C6\}$ by Lemma~\ref{lem:conditionrestrictions} for every $l \in \{f,g\}$, and $X \neq C7$ by Lemmas~\ref{lem:AtMostOnePullFromTunnel}\ref{enum:oneway} and~\ref{lem:ExitPairsDominateTunnelInside}, so that $X = C2$. Since minor 3-faces whose middle $C$-edge is incident to a thin minor 2-face are not contained in any tunnel, we have $l = g$. By $C2$, $e$ is a $C$-edge of a 2-arch of $f$, which contradicts Lemma~\ref{lem:AtMostOnePullFromTunnel}\ref{enum:2archOnTrack} or \ref{enum:no2arch}. We conclude that $Y \neq C7$, which implies $X \neq C7$. We distinguish the remaining options for $X$ and $Y$ in $\{C2,\dots,C6\}$.

Assume $X = C2$. If $m_f = 3$, $e$ is an extremal $C$-edge of a 3-face, so that Lemma~\ref{lem:conditionrestrictions} and planarity imply $Y \notin \{C3,C4,C5,C6\}$ for every $l \in \{f,g\}$. Hence, $m_f = 2$. Then $l = g$, as the remaining options $Y \in \{C3,C4,C5,C6\}$ for $l = f$ require $m_f \geq 3$.
By Lemma~\ref{lem:conditionrestrictions} (for $C2$ and $C6$), $Y \notin \{C2,C3,C6\}$.
In the remaining case, $Y \in \{C4,C5\}$ contradicts $m_f = 2$.

Assume $X = C3$. Then $e$ is the middle $C$-edge of a 3-arch $A$ of $f$, so that $l = g$ implies $Y \notin \{C3,C4,C5,C6\}$ by Lemma~\ref{lem:conditionrestrictions}. We conclude $l = f$.
If $Y \in \{C4,C5\}$, $g$ is the only opposite face of $A$, which implies by $C3$ that $f$ is a 3-face; this contradicts that $f$ has a 4-arch.
Thus $Y = C6$, which contradicts Lemma~\ref{lem:conditionrestrictions} (for $C6$).

Assume $X \in \{C4,C5\}$. If $l = f$, $Y \notin \{C5,C6\}$, as 2-faces do not have 3-arches, so let $l = g$.
Then $Y \in \{C4,C5\}$ contradicts Lemma~\ref{lem:conditionrestrictions} (for $Y$) and $Y = C6$ contradicts planarity.

Assume $X = C6$. Then $Y = C6$ and thus $l = g$, which contradicts that $G$ is plane.
\end{proof}
}

By Lemma~\ref{lem:atMost1OverAnyEdge}, we know that whenever weight~1 is pulled over an edge $e$ by some condition~$C1$--$C7$, no other condition is satisfied on $e$, so that~1 is the final amount of weight transferred over $e$.

\subsection{The Proof}
Throughout this section, let $w$ denote the weight function on the set of faces of $H$ after our discharging rule has been applied. Clearly, $\sum_{f \in F(H)} w(f) = 2c$ still holds. In order to prove Inequality~\eqref{eq:inequality1} if $V^- \neq \emptyset$ and otherwise Inequality~\eqref{eq:inequality2}, we aim to show that every minor face $f$ satisfies $w(f) \geq 4$ if $f$ is thick and $w(f) \geq 2$ if $f$ is thin such that no face has negative weight.

\longversion{For $S \subseteq E(C)$ and the set $X$ of $C$-edges of a face $f$ of $H$, let the (weight) \emph{contribution} of $S$ to $f$ be $|S \cap X|$ (i.e.\ the initial weight the edges of $S$ give to $w(f)$) plus the sum of all weights pulled by $f$ over edges in $S \cap X$ minus the sum of all weights pulled by opposite faces of $f$ over edges in $S \cap X$. The \emph{contribution} of an arch $A$ to $f$ is the contribution of its $C$-edges to $f$; this way, every proper arch $A$ contributes weight $w(f(A))$ to $f(A)$. Since $f$ looses weight at most~1 over every of its $C$-edges by Lemma~\ref{lem:atMost1OverAnyEdge}, we have $w(f) \geq x$ if a set $S$ contributes weight $x$ to $f$.}

By our discharging rule, most pulls occur on $C$-edges that are extremal or adjacent to one; the following definition captures the remaining pulls and will be used in a final counting argument. Let \pulls{g}{e}{X} be a \emph{mono-pull} if $X = C3$ and $e$ and its two adjacent edges in $C$ are incident to a common face $f \neq g$. \longversion{An edge $e \in E(C)$ is \emph{mono} if it is incident to a face $g$ such that \pulls{g}{e}{C3} is satisfied and a mono-pull, and \emph{non-mono} otherwise.}

\begin{lemma}\label{lem:notAdjacent}
For two $C$-edges $e$ and $b$ of a minor face $f$, let \pulls{g}{e}{X} and \pulls{h}{b}{Y} such that $f \notin \{g,h\}$ and $X$ and $Y$ are not contained in $\{C2,C7\}$. Then
\begin{enumerate}
	\item if \pulls{g}{e}{X} is no mono-pull, $e$ is either an extremal $C$-edge of $f$ or adjacent to one (more precisely, an extremal $C$-edge of $f$ if and only if $X \in \{C3,C6\}$), and\label{enum:extremal}
	\item $e$ and $b$ have distance at least three in $C$.\label{enum:notAdjacent}
\end{enumerate}
\end{lemma}
\begin{proof}
Consider Claim~\ref{enum:extremal}. %
Since $f$ is minor, $X \notin \{C1,C2,C7\}$. If $X = C3$, $e$ is an extremal $C$-edge of $f$, since \pulls{g}{e}{X} is no mono-pull. For every $X \in \{C4,C5\}$, $e$ is adjacent to an extremal $C$-edge of $f$ by definition of~$X$ (for $X = C5$, this follows from the transfer pair). If $X = C6$, $e$ is an extremal $C$-edge of $f$.

For Claim~\ref{enum:notAdjacent}, assume to the contrary that $e$ and $b$ have distance at most two in $C$. Since $f$ is minor, $C1 \notin \{X,Y\}$. Let $X = C3$ and let $B$ be the 3-arch of $g$ that has middle $C$-edge $e$. Then the existence of $B$ and planarity imply $Y \notin \{C3,C4,C5\}$, and Lemma~\ref{lem:conditionrestrictions} and planarity imply $Y \neq C6$. Hence, $X \in \{C4,C5,C6\}$ and, by symmetry, the same holds for $Y$. Then $Y \notin \{C4,C5,C6\}$ by planarity, Lemma~\ref{lem:conditionrestrictions} and the respective condition $Y$ imposes on $f$. This is a contradiction.
\end{proof}

\longversion{
\begin{lemma}\label{lem:pedestalCost}
Let \pulls{g}{e}{X}, $f$ be the $e$-opposite face of $g$, $B$ be the arch of $g$ shown in Figure~\ref{fig:Conditions} (for $X \in \{C2,C6\}$, let $B$ be the proper arch of $g$), and $S$ be the set of common $C$-edges of $f$ and $B$.
\begin{itemize}
	\item If $X = C3$ and $|S| = 3$ (i.e.\ \pulls{g}{e}{X} is a mono-pull), every of the two extremal $C$-edges of $B$ contributes weight at least $1$ to $f$. %
	\item If $X \in \{C4,C5\}$, $S$ contributes weight at least $2$ to $f$.
	\item If $X = C6$, $S$ contributes weight at least $1$ to $f$.
\end{itemize}
\end{lemma}
\begin{proof}
For every $X \in \{C3,\dots,C6\}$, $f$ is minor. Assume $X = C3$ and $|S| = 3$. Then every $C$-edge of $B$ is incident to $f$, which implies $m_g = 3$ (by $C3$). Assume to the contrary that $S$ contributes weight at most~$1$ to $f$. Then \pulls{g}{b}{Y} for a $C$-edge $b \neq e$ of $B$ and some condition~$Y$. Since $f$ is minor, $Y \neq C1$. By $m_g \geq 3$ and the definition of~$C2$ in that case, $Y \neq C2$. By Lemma~\ref{lem:notAdjacent}\ref{enum:notAdjacent}, $Y \notin \{C3,C4,C5,C6\}$. Hence, $Y = C7$. Then $(g,b)$ is a transfer pair, which contradicts that every $C$-edge of $B$ is incident to $f$.

Let $X \in \{C4,C5,C6\}$. Then $m_g \geq 4$, $|S| = 3$ if $X \in \{C4,C5\}$, and $|S|=2$ if $X = C6$. Assume to the contrary that $g$ pulls weight over an edge $b \neq e$ of $S$ by some condition~$Y$. Since $f$ is minor, $Y \neq C1$. Since $m_g \geq 4$, $Y \neq C2$. By Lemma~\ref{lem:notAdjacent}\ref{enum:notAdjacent}, $Y \notin \{C3,C4,C5,C6\}$. Hence, $Y = C7$. Then $(g,b)$ is a transfer pair, which contradicts $|S|=3$ if $X \in \{C4,C5\}$ and Lemma~\ref{lem:conditionrestrictions} if $X = C6$.
\end{proof}

\begin{lemma}\label{lem:smallestkArch}
For a minor face $f$, let $A$ be an arch of $f$ with minimal $m_A$ such that a face $h \neq f$ pulls weight over a $C$-edge $b$ of $A$ by Condition $Y \in \{C2,C7\}$. Then $w(f) \geq 2$ and, if $f$ is thick, $w(f) \geq 4$.
\end{lemma}
\begin{proof}
Assume that $w(f) < 4$, as otherwise the claim holds. Then $w(f) \leq 3$ by Lemma~\ref{lem:atMost1OverAnyEdge} and, by $C1$, at most one $C$-edge of $f$ is incident to a major face (we will use this throughout the proof). Let $A_h$ be the proper arch of $h$ if $Y = C2$ and the unique 3-arch of $h$ that has $C$-edge $b$ if $Y = C7$. In both cases, $h$ is thick (for $Y = C7$, as $A_h$ is a transfer arch) and minor and $b$ is an extremal $C$-edge of $A_h$. If $Y = C7$, $(h,b)$ is a transfer pair. In that case, let $T$ be the tunnel track containing $A_h$ whose exit pair is on-track with $(h,b)$; by Lemma~\ref{lem:ExitPairsDominateTunnelInside} and \pulls{h}{b}{Y}, the exit pair of $T$ satisfies $C2$ such that the exit face is a 2-face.

Assume to the contrary that $m_{A,A_h} \geq 2$. If $Y = C2$, this implies $m_h = 2 = m_{A,h}$. Since $G$ has minimum degree at least three, the middle $C$-vertex of $h$ is then an extremal $C$-vertex of an arch of $A$, which contradicts the minimality of $m_A$. Hence, $Y = C7$. Then $m_{A,A_h} = 2$, as $m_{A,A_h} \neq 3$ by the definition of transfer arches. Consider the non-extremal $C$-vertex $v$ of $A_h$ that is incident to $b$. By minimality of $m_A$ and $\deg_G(v) \geq 3$, $v$ is an extremal $C$-vertex of a 2-arch of $A_h$, which contradicts Lemma~\ref{lem:AtMostOnePullFromTunnel}\ref{enum:2archOnTrack}. We conclude that $m_{A,A_h} = 1$; in particular, $b$ is an extremal $C$-edge of $A$.

This shows that no opposite face of $f$ pulls weight over a non-extremal $C$-edge of $A$ using $C2$ or $C7$. We distinguish the following cases for $m_A$.

\begin{description}[style=nextline]
	\item[Case $m_A = 2$:]
	Then $Y = C2$ contradicts Lemma~\ref{lem:conditionrestrictions}, and $Y = C7$ contradicts Lemma~\ref{lem:AtMostOnePullFromTunnel}\ref{enum:2archOnTrack} or~\ref{enum:no2arch} (which one depends on whether there is another arch of $T$ that has $C$-edge $b$).

	\item[Case $m_A = 3$:]
	Let $v_2v_3$ be the extremal $C$-edge of $A$ that is different from $b$, and let $g$ be the (possibly major) $v_2v_3$-opposite face of $f$, as shown in Figure~\ref{fig:Smallest3Arch5}. If $Y = C2$, $v_1$ is an extremal $C$-vertex of $h$. If $Y = C7$, $(h,b)$ is a transfer pair, so that $v_1$ or $v_2$ is an extremal $C$-vertex of $h$ and $v_{-2}v_{-1}$ is not incident to $f$. Hence, in all cases, either $v_1$ or $v_2$ is an extremal $C$-vertex of $h$, which implies $g \neq h$.
	
	By minimality of $m_A$, $v_0v_2 \notin E(G)$. Since $\deg_G(v_2) \geq 3$ and not both $v_1v_2$ and $v_2v_3$ are incident to major faces, $v_2$ is an extremal $C$-vertex of an arch whose face is not $f$. Let $B$ be such an arch with minimal $m_B$. Since $H$ has no minor 1-face, $f(B) \in \{g,h\}$. By the existence of $B$ and planarity, $v_1v_2$ is not incident to a thin minor 2-face, so that $A$ is contained in a tunnel. Let $T$ be the track of this tunnel whose exit pair is on-track with $(h,b)$; note this is consistent with the definition of $T$ in the case $Y = C7$, since $A$ and $A_h$ are consecutive and thus in the same tunnel track. By Lemma~\ref{lem:ExitPairsDominateTunnelInside} and \pulls{h}{b}{Y}, the exit pair of $T$ satisfies $C2$. Thus, $m_B = 2$ contradicts Lemma~\ref{lem:AtMostOnePullFromTunnel}\ref{enum:2archOnTrack} or~\ref{enum:no2arch}. We conclude that $m_B \geq 3$.
	
		We show next that the weight contribution of $\{v_{-1}v_0,v_1v_2\}$ to $f$ is at least one and, if $v_1v_2$ is not incident to $g$, at least two. By Lemma~\ref{lem:conditionrestrictions}, planarity and the existence of $B$, the $v_1v_2$-opposite face of $f$ does not pull weight over $v_1v_2$ by any condition. This gives the first claim, so assume that $v_1v_2$ is not incident to $g$. If $v_1v_2$ is incident to a major face, the second claim follows straight from \pulls{f}{v_1v_2}{C1}. Otherwise, $v_1v_2$ is incident to $h \neq g$, as $H$ has no minor 1-face. Then $Y = C7$, $A_h$ is non-proper and $(h,b)$ is a transfer pair, so that $v_{-2}v_{-1}$ is not incident to $f$ and $v_{-1}v_0$ is either incident to a major face and $g$ is minor or incident to $f$. If $v_{-1}v_0$ is incident to a major face and $g$ is minor, the second claim follows from \pulls{f}{v_1v_2}{C3}; hence, $v_{-1}v_0$ is incident to $f$. Then $h$ does not pull weight over $v_{-1}v_0$ using Condition~$C3$, as $v_1v_2$ is incident to $h$. By Lemma~\ref{lem:conditionrestrictions}, planarity and the existence of $B$, $h$ does not pull weight over $v_{-1}v_0$ by any other condition, which gives the second claim.
		
		If $g$ is major, \pulls{f}{v_2v_3}{C1} and $v_1v_2$ is not incident to $g$, which contradicts $w(f) \leq 3$ by the previous result; hence, $g$ is minor. If $f(B) = g$, $m_B \geq 3$ implies that $v_4v_5$ is incident to $g$. If $f(B) = h$, $v_1v_2$ is incident to $h$, so that $Y = C7$ and $m_B \geq m_{A_h}+1 \geq 4$; then the minimality of $m_B$ implies $m_g \geq 4$, as $g$ is minor. Assume that $f$ is thin; then $m_f = m_A = 3$. If $v_1v_2$ is not incident to $g$, the previously proven weight contribution of $\{v_{-1}v_0,v_1v_2\}$ gives the claim $w(f) \geq 2$. Otherwise, $v_1v_2$ is incident to $g$ and, by the existence of $B$, $v_1v_2$ is not an extremal $C$-edge of a 3-arch. Then \pulls{f}{v_1v_2}{C3} gives the claim.
		
		Hence, in the remaining case, $f$ is thick, $g$ is minor, $m_g \geq 3$ and $v_4v_5$ is incident to $g$. We distinguish the following cases for $T$.
	
\begin{figure}[!htb]
	\centering
	\subcaptionbox{$m_A = 3$ when $Y = C7$.
			\label{fig:Smallest3Arch5}}%
			{\includegraphics[page=1,scale=0.9]{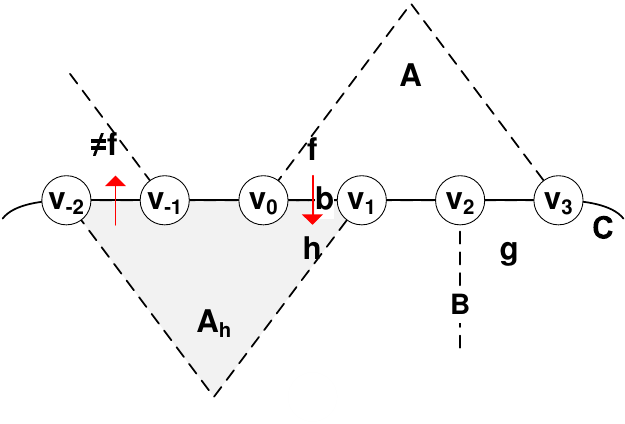}}
	\hspace{0.3cm}
	\subcaptionbox{$A$ is a transfer arch of $T$ and $Y = C2$.
			\label{fig:Smallest3Arch5aC2}}%
			{\includegraphics[page=1,scale=0.9]{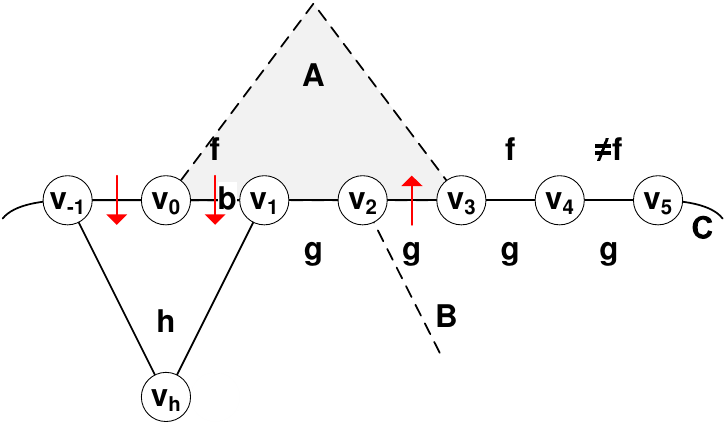}}
	\hspace{0.3cm}
	\subcaptionbox{$A$ is a transfer arch of $T$ and $Y = C7$.
			\label{fig:Smallest3Arch5a}}%
			{\includegraphics[page=1,scale=0.9]{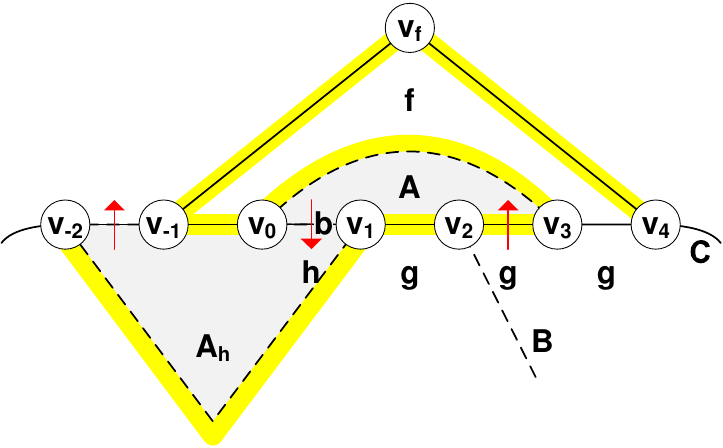}}
	\hspace{0.3cm}
	\subcaptionbox{$A$ is thick and not a transfer arch of $T$, \pulls{g}{v_3v_4}{C7} and $Y = C7$.
			\label{fig:Smallest3Arch5b}}%
			{\includegraphics[page=1,scale=0.9]{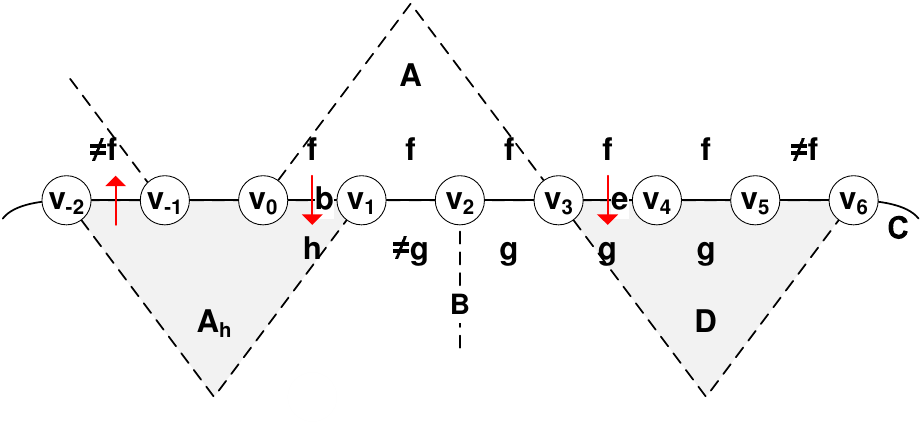}}
	\hspace{0.3cm}
	\caption{Case $m_A = 3$.
	}\label{fig:Smallest3Arches}
\end{figure}

	\begin{description}[style=nextline]
		\item[$A$ is a transfer arch of $T$:]
	Then $(f,v_2v_3)$ is a transfer pair of $T$. Since the exit pair of $T$ satisfies $C2$, we have \pulls{f}{v_2v_3}{C7} for both $Y = C2$ and $Y = C7$. If $v_1v_2$ is not incident to $g$, the weight contribution of $\{v_{-1}v_0,v_1v_2\}$ to $f$ is at least two by the previous result, which contradicts $w(f) \leq 3$. Hence, $v_1v_2$ is incident to $g$, which implies $f(B) = g$ (see Figures~\ref{fig:Smallest3Arch5aC2} and~\ref{fig:Smallest3Arch5a}). We do not have \pulls{f}{v_1v_2}{C3}, as this contradicts $w(f) \leq 3$. Since all other requirements of $C3$ are satisfied for this pull (in particular, $v_1v_2$ is not an extremal $C$-edge of a 3-arch by planarity), $v_3v_4$ is incident to $f$. Since $(f,v_2v_3)$ is a transfer pair, this implies that $v_4$ is an extremal $C$-vertex of $f$.
	
	Assume that $Y = C2$ (see Figure~\ref{fig:Smallest3Arch5aC2} for the case $m_h = 2$).
	Since $\{v_{-1}v_0,v_1v_2\}$ contributes weight at least one to $f$ and $w(f) \leq 3$, we have \pulls{g}{v_3v_4}{X} for some condition~$X$. As $f$ is thick and minor and $m_g \geq 3$, $X \notin \{C1,C2\}$. Since $f$ is incident to $v_3v_4$ and $v_1v_2$ is incident to $g$, $X \notin \{C3,C4,C5,C6\}$, so that $X = C7$. By planarity and the existence of $B$, $(g,v_3v_4)$ is then a transfer pair and $v_3$ is an extremal $C$-vertex of a transfer arch, which contradicts that $v_1v_2$ is incident to $g$.
	
	Hence, $Y = C7$. Then, since $v_{-2}v_{-1}$ is not incident to $f$, either $v_{-1}$ or $v_0$ is an extremal $C$-vertex of $f$. By Lemma~\ref{lem:AtMostOnePullFromTunnel}\ref{enum:no4arch}, $v_{-1}$ is an extremal $C$-vertex of $f$. Then $C$ is extendable by Figure~\ref{fig:Smallest3Arch5a}.

	\item[$A$ is not a transfer arch of $T$:]
	Then $(f,v_2v_3)$ is not a transfer pair of $T$, which implies that $v_3v_4$ is incident to $f$ and that, if $v_1v_2$ is incident to $g$, $v_4v_5$ is incident to $f$ (as $f$ is thick, $g$ is minor, $m_g \geq 3$ and $g \neq h$).

	We show that the weight contribution of $S := \{v_2v_3,v_3v_4,v_4v_5\}$ to $f$ is at least two and, if $v_1v_2$ is incident to $g$, at least three. This contradicts $w(f) \leq 3$, as the weight contribution of $\{v_{-1}v_0,v_1v_2\}$ to $f$ is at least one and, if $v_1v_2$ is not incident to $g$, at least two. Let $e := v_iv_{i+1}$ be an edge of $S$ such that $e$ is incident to $f$ and some condition $X$ satisfies \pulls{g}{e}{X}; we may assume that $e$ exists, as otherwise $S$ satisfies the claim, since $v_4v_5$ is incident to $f$ if $v_1v_2$ is incident to $g$ by the result above.
	
	Since $f$ is thick and minor and $m_g \geq 3$, $X \notin \{C1,C2\}$. Assume $X = C3$. By planarity and $m_B \geq 3$, $e \neq v_2v_3$. Since $e$ and the edge $v_{i-1}v_i$ are incident to $f$, Condition~$C3$ implies that $v_{i-2}v_{i-1}$ is not incident to $g$. As $v_2v_3$ is incident to $g$, we have $e = v_3v_4$ such that $v_4v_5$ is incident to a minor face $p \notin \{f,g\}$; in particular, $p$ is neither major nor $f$. Since $(f,v_2v_3)$ is not a transfer pair of $T$, then $v_1v_2$ is neither incident to a major face nor to $g$.
	Since $H$ has no minor 1-face, $v_1v_2$ is incident to $h$, so that $Y = C7$ and either $v_{-1}$ or $v_0$ is an extremal $C$-vertex of $f$. By Lemma~\ref{lem:AtMostOnePullFromTunnel}\ref{enum:no4arch}, $v_{-1}$ is an extremal $C$-vertex of $f$. Then $C$ is extendable by the replacement as shown in Figure~\ref{fig:Smallest3Arch5a}.
	
	Assume that $X \in \{C4,C5\}$. Since $m_B \geq 3$ and $v_2v_3$ is incident to $g$, we have $e = v_3v_4$ such that $v_1v_2$ is not incident to $g$ and $v_4v_5$ is incident to $f$. Then $S$ contributes weight at least two to $f$ by Lemma~\ref{lem:pedestalCost}, as claimed. By the existence of $B$, Lemma~\ref{lem:conditionrestrictions}, and the fact that $v_3v_4$ is incident to $f$ and $v_2v_3$ to $g$, we have $X \neq C6$.
	
	We conclude that $X = C7$, so that $(g,e)$ is a transfer pair. Let $D$ be the transfer arch that has extremal $C$-edge $e$ such that $f(D) = g$. Since $A$ is not a transfer arch of $T$, $D$ is not a transfer arch of $T$. By Lemmas~\ref{lem:AtMostOnePullFromTunnel}\ref{enum:oneway} and~\ref{lem:ExitPairsDominateTunnelInside} and \pulls{h}{b}{Y}, $D$ is not in the same tunnel as $T$. In particular, $D$ is not consecutive to $A$, so that $e \neq v_2v_3$ and $D$ has $C$-edge $v_5v_6$. Since $(g,e)$ is a transfer pair and $v_2v_3$ is incident to $g$, we have $e = v_3v_4$ such that $v_1v_2$ is not incident to $g$ and $v_5v_6$ is not incident to $f$ (see Figure~\ref{fig:Smallest3Arch5b}); for the same reason, $v_4v_5$ is incident to $f$, as $f$ has no 3-arch that is consecutive to $D$. Since we excluded all other options for $e$ and $X$, $\{v_2v_3,v_4v_5\} \subseteq S$ contributes weight at least two to $f$. This gives the claim, as $v_1v_2$ is not incident to $g$.
	\end{description}

	\item[Case $m_A = 4$:]
	Let $g$ be the (possibly major) $v_2v_3$-opposite face of $f$ in Figure~\ref{fig:Smallest4Arch}. Since $h$ is thick, $g$ is thick. As argued at the very beginning of case $m_A = 3$, either $v_1$ or $v_2$ is an extremal $C$-vertex of $h$, which implies $g \neq h$.
	
	We first show that $\{v_1v_2\}$ contributes weight at least one to $f$ and, if $v_1v_2$ is incident to $h$, $\{v_{-1}v_0,v_1v_2\}$ contributes weight at least two to $f$. By Lemma~\ref{lem:conditionrestrictions}, planarity, $g \neq h$ and minimality of $m_A$, the $v_1v_2$-opposite face of $f$ does not pull weight over $v_1v_2$ by any condition. This gives the first claim, so assume that $v_1v_2$ is incident to $h$. Then $Y = C7$ and $(h,b)$ is a transfer pair. This implies that $v_{-1}v_0$ is incident to $f$, as $A$ has no 3-arch with extremal $C$-vertex~$v_0$ by minimality of $m_A$. Then $v_{-1}$ is an extremal $C$-vertex of $f$ and $h$ does not pull weight over $v_{-1}v_0$ using $C3$. By Lemma~\ref{lem:conditionrestrictions} and $g \neq h$, $h$ does not pull weight over $v_{-1}v_0$ using any other condition. This gives the claim.
		
		Assume to the contrary that $g$ is major. Then \pulls{f}{v_2v_3}{C1} and $v_1v_2$ is incident to a minor face. Since $H$ has no minor 1-face, $v_1v_2$ is incident to $h$, which contradicts $w(f) \leq 3$ by the claim just proven. Hence, $g$ is thick and minor. If $v_3v_4$ is not incident to $g$, $g$ is a thick minor 2-face, as $h \neq g$ and $H$ has no minor 1-face. This contradicts that no opposite face of $f$ pulls weight over a non-extremal $C$-edge of $A$ using $C2$ or $C7$ (which we proved by minimality of $m_A$). We conclude that $g \neq h$ is thick and minor and incident to $v_3v_4$.
		
\begin{figure}[!htb]
	\centering
	\subcaptionbox{$Y = C2$ and $g$ has a 3-arch $B$ with middle $C$-edge $v_3v_4$. %
			\label{fig:Smallest4Arch}}[0.67\linewidth]
			{\includegraphics[page=1,scale=0.9]{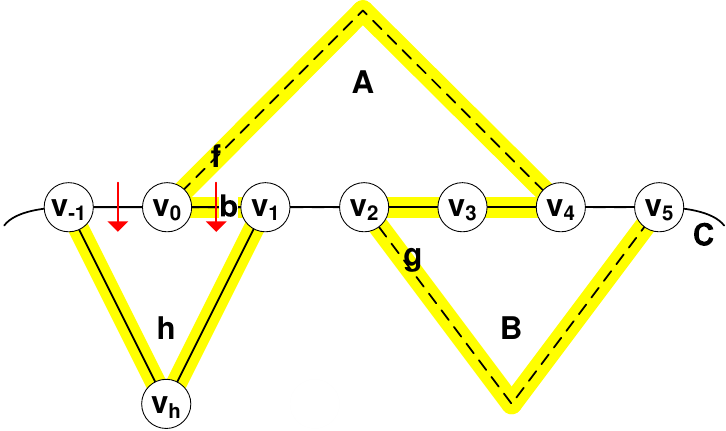}}
	\hspace{0.3cm}
	\subcaptionbox{$Y = C7$ and \pulls{g}{v_3v_4}{C7}.
			\label{fig:Smallest4ArchC7C7}}%
			{\includegraphics[page=1,scale=0.9]{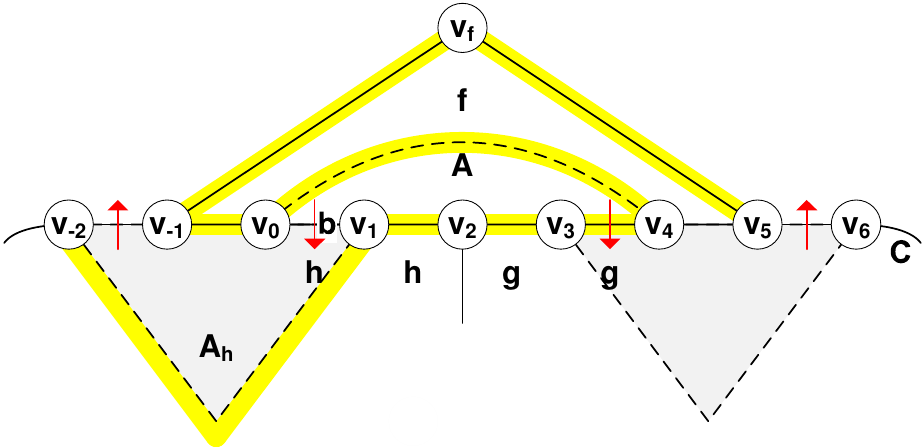}}
	\hspace{0.3cm}
	\subcaptionbox{$Y = C2$, \pulls{g}{v_3v_4}{C5} and $v_3v_5 \in E(G)$.
			\label{fig:Smallest4ArchC4C5}}%
			{\includegraphics[page=1,scale=0.9]{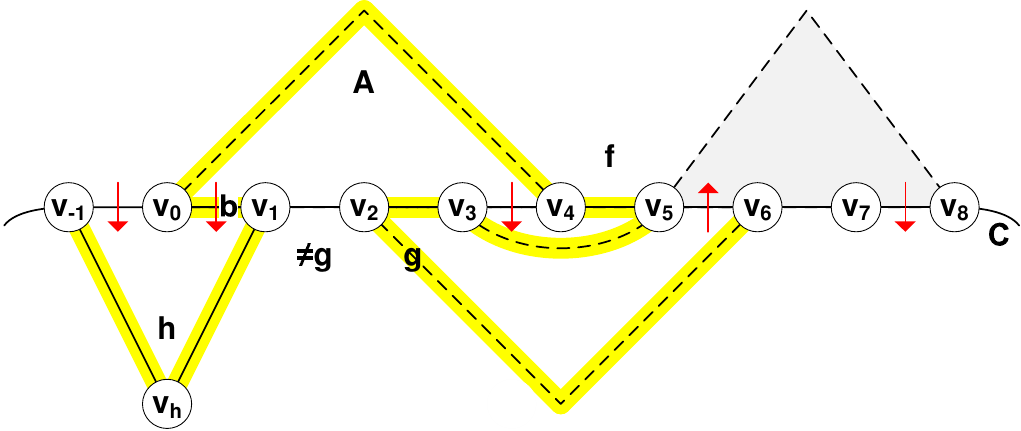}}
	\hspace{0.3cm}
	\caption{Case $m_A = 4$.
	}\label{fig:smallest4arches}
\end{figure}

		Assume to the contrary that an opposite face of $f$ pulls weight over a $C$-edge of $A$ different from $b$ using $C2$ or $C7$. Since this is impossible for non-extremal $C$-edges of $A$, this edge is $v_3v_4$ and we have (by $C2$ and the fact that $g$ is incident to $v_2v_3$) \pulls{g}{v_3v_4}{C7}. By symmetry and the result about the contribution $\{v_{-1}v_0,v_1v_2\}$, then $v_1v_2$ is not incident to a major face. Since $(g,v_3v_4)$ is a transfer pair, $v_1v_2$ is incident to $h$ and thus $Y = C7$. By minimality of $m_A$, neither $v_0$ nor $v_4$ is an extremal $C$-vertex of a 3-arch of $A$. Since $(h,b)$ and $(g,v_3v_4)$ are transfer pairs, this implies that $f$ has extremal $C$-vertices $v_{-1}$ and $v_5$. Then $f$ is thick and $C$ extendable by Figure~\ref{fig:Smallest4ArchC7C7}. We conclude that $b$ is the only $C$-edge of $A$ over which an opposite face of $f$ pulls weight using $C2$ or $C7$.
		
		Assume to the contrary that $g$ has a 3-arch $B$ with middle $C$-edge $v_3v_4$. Then $C$ is extendable by Figure~\ref{fig:Smallest4Arch} (and an analogous replacement for $m_h = 3$) if $Y = C2$, and by appending $v_1v_0Av_4v_3v_2v_5$ to the tunnel replacement of Lemma~\ref{lem:AtMostOnePullFromTunnel}\ref{enum:2archOnTrack} if $Y = C7$ (this adds at most $3+n_5(G)$ new vertices to $C$, as $A$ and the 3-arch of $g$ add at most one each).

		Assume to the contrary that $A$ has a $C$-edge $e \neq b$ such that the $e$-opposite face of $f$ pulls weight over $e$ by some condition~$X \notin \{C2,C7\}$. Since $f$ is minor, $X \neq C1$. If $X = C3$, we have $e = v_3v_4$ by Lemma~\ref{lem:conditionrestrictions} and planarity, which contradicts that $g$ has no 3-arch with middle $C$-edge $v_3v_4$. Assume $X \in \{C4,C5\}$. By Lemma~\ref{lem:conditionrestrictions}, $h \neq g$ and planarity, we have $e = v_3v_4$ and $f$ is thick. Since $A$ has no 3-arch with extremal $C$-vertex $v_0$, $G$ contains $v_1v_3$ or $v_3v_5$ by Lemma~\ref{lem:conditionrestrictions}. In the latter case, $C$ is extendable by Figure~\ref{fig:Smallest4ArchC4C5} if $Y = C2$, and by appending $v_1v_0Av_4v_5v_3v_2v_6$ to the tunnel replacement of Lemma~\ref{lem:AtMostOnePullFromTunnel}\ref{enum:2archOnTrack} if $Y = C7$. In the former case, $C$ is extendable by the replacement $v_0Av_4v_3v_1v_2v_6v_5v_7$ if $X = C4$, and by prepending $v_0Av_4v_3v_1v_2v_6v_5$ to the tunnel replacement if $X = C5$. By Lemma~\ref{lem:conditionrestrictions}, planarity, $\deg_G(v_3) \geq 3$ and the fact that $A$ has no 3-arch with extremal $C$-vertex~$v_0$, we have $X \neq C6$. Hence, every edge in $\{v_1v_2,v_2v_3,v_3v_4\}$ contributes weight at least one to $f$.

		This gives $w(f) \geq 2$ if $f$ is thin, so let $f$ be thick. If $v_1v_2$ is incident to $h$, the contribution of $\{v_{-1}v_0,v_1v_2\}$ to $f$ of weight at least two contradicts $w(f) \leq 3$. If $v_1v_2$ is incident to a major face, \pulls{f}{v_1v_2}{C1} contradicts $w(f) \leq 3$. Hence, $v_1v_2$ is incident to $g$ and so is $v_3v_4$, which gives $m_g \geq 4$, as $\{v_1,v_4\}$ is not a 2-separator of $G$. Assume to the contrary that $v_4$ is an extremal $C$-vertex of $f$. If $Y = C2$, \pulls{f}{v_2v_3}{C4} contradicts $w(f) \leq 3$; hence, let $Y = C7$. Then $v_0$ is not an extremal $C$-vertex of a 2-arch of $h$ or $f$ by Lemma~\ref{lem:AtMostOnePullFromTunnel}\ref{enum:2archOnTrack} or~\ref{enum:no2arch}, $(h,b)$ is a transfer pair, and $v_3v_4$ is not the middle $C$-edge of a 3-arch of $g$. This implies \pulls{f}{v_2v_3}{C5}, which contradicts $w(f) \leq 3$. We conclude that $v_4v_5$ is incident to $f$ and $g$.
		
		We show that $\{v_4v_5\}$ contributes weight at least one to $f$, which contradicts $w(f) \leq 3$. Assume to the contrary that $g$ pulls weight over $v_4v_5$ by some Condition~$X$. As $f$ is minor and $m_g \geq 4$, $X \notin \{C1,C2\}$. Since $m_g \geq 4$, and $v_3v_4$ is incident to $f$ and $v_2v_3$ to $g$, we have $X \neq C3$. By planarity and the facts that $v_2v_3$ is incident to $g$ and $A$ has no 3-arch with extremal $C$-vertex $v_0$, $X \notin \{C4,C5,C6\}$. Hence, $X = C7$, so that $(g,v_4v_5)$ is a transfer pair, which contradicts that $v_2v_3$ is incident to $g$.

	\item[Case $m_A \geq 5$:]
		Let $b = v_0v_1$ such that $v_0$ is an extremal $C$-vertex of $A$, let $z := m_A$ and let $S$ be the set of $C$-edges of $A$ that are different from $b$. By precisely the same arguments as used at the beginning of Case~$m_A = 4$, $v_2v_3$ is not incident to $h$, $\{v_1v_2\}$ contributes weight at least one to $f$ and, if $v_1v_2$ is incident to $h$, $\{v_{-1}v_0,v_1v_2\}$ contributes weight at least two to $f$. Since $w(f) \leq 3 < 4 \leq |S|$, $S$ contains an edge that contributes weight zero to $f$.
		
		Assume that $S$ contains a non-mono edge $e$ that contributes weight zero to $f$. Since $\{v_1v_2\}$ contributes weight at least one to $f$, $e \neq v_1v_2$. Let $X$ be the condition by which the $e$-opposite face $p$ of $f$ pulls weight over $e$. Since $f$ is minor, $X \neq C1$. Assume $X \in \{C2,C7\}$. Then $e = v_{z-1}v_z$, as no opposite face of $f$ pulls weight over a non-extremal $C$-edge of $A$ using $C2$ or $C7$. By applying the symmetric version of the statement about the contribution of $\{v_1v_2\}$ above, $\{v_{z-2}v_{z-1}\}$ contributes weight at least one to $f$ and, if $v_{z-2}v_{z-1}$ is incident to $p$, $\{v_{z-2}v_{z-1},v_zv_{z+1}\}$ contributes weight at least two to $f$. In particular, $v_{z-2}v_{z-1}$ and $e$ are non-mono (by $C3$, only a minor 3-face and thus not $f$ may pull weight over a mono edge). Assume $X \in \{C3,C4,C5,C6\}$. Since $e$ is non-mono, Lemma~\ref{lem:notAdjacent}\ref{enum:extremal} and $e \notin \{b,v_1v_2\}$ imply $e \in \{v_{z-2}v_{z-1},\allowbreak v_{z-1}v_{z}\}$. By Lemma~\ref{lem:notAdjacent}\ref{enum:notAdjacent} and the result for $X \in \{C2,C7\}$, $\{v_{z-2}v_{z-1},\allowbreak v_{z-1}v_{z}\}$ contributes weight at least one to $f$ and no edge of $\{v_{z-3}v_{z-2},v_{z-2}v_{z-1},\allowbreak v_{z-1}v_{z}\}$ is mono. We conclude in all cases that
		\begin{itemize}
			\item $e$ is the only non-mono edge of $S$ that contributes weight zero to $f$,
			\item $e \in \{v_{z-2}v_{z-1},\allowbreak v_{z-1}v_z\}$,
			\item $v_{z-2}v_{z-1}$ and $v_{z-1}v_z$ are non-mono and, if $X \notin \{C2,C7\}$, $v_{z-3}v_{z-2}$ is non-mono, and
			\item the weight contribution of $\{v_{z-2}v_{z-1},\allowbreak v_{z-1}v_z,v_zv_{z+1}\}$ to $f$ is at least one and, if $X \in \{C2,C7\}$ and $v_{z-2}v_{z-1}$ is incident to $p$, at least two.
		\end{itemize}
		
		Assume now that $S$ contains a mono edge $e'$ and let $p'$ be the $e'$-opposite face of $f$. Since $\{v_1v_2\}$ contributes weight at least one to $f$, $e' \neq v_1v_2$. By $m_f > 3$, we have \pulls{p'}{e'}{C3}, so that $p'$ is a minor 3-face by definition of~$C3$. Since $h$ is thick, $p'$ is thick. By Lemma~\ref{lem:pedestalCost}, every of the two extremal $C$-edges of $p'$ contributes weight at least one to $f$; by Lemma~\ref{lem:notAdjacent}\ref{enum:notAdjacent}, every two mono edges in $S$ have distance at least three in $C$. Since $w(f) \leq 3$, we thus conclude that $e'$ is the only mono edge of $S$.
		
		We conclude that $S$ contains at most two edges that contribute weight zero to $f$, namely the non-mono edge $e$ and the mono edge $e'$ above. Since \pulls{h}{b}{Y}, $w(f) \geq m_A-3$. Since $w(f) \leq 3$, this implies $5 \leq m_A \leq 6$ and $w(f) \geq 2$. If $f$ is thin, this gives the claim, so assume that $f$ is thick. We distinguish the following cases.
		
	\begin{description}[style=nextline]
		\item[Case $m_A = 5$:]
		First, assume that $e'$ exists. If $e' = v_2v_3$, Lemma~\ref{lem:conditionrestrictions} implies $v_2v_5 \notin E(G)$, as $p'$ is a thick minor 3-face. Since $\{v_1,v_4\}$ is not a 2-separator of $G$ and $v_0$ is not an extremal $C$-vertex of any arch of $A$ except $A$ itself, $G$ contains $v_3v_5$ and $v_2v_4$. Then $C$ is extendable by Figure~\ref{fig:Smallest5ArchMono1}. Hence, $e' = v_3v_4$. Then the results about non-mono edges imply that $e$ does not exist, so that $w(f) = 3$. If $v_1v_2$ is incident to $h$, the contribution of $\{v_{-1}v_0,v_1v_2\}$ to $f$ contradicts $w(f) \leq 3$. Otherwise, $v_1v_2$ is incident to a major face, since $H$ has no minor 1-face. This contradicts $w(f) \leq 3$.
		
\begin{figure}[!htb]
	\centering
	\subcaptionbox{$m_A = 5$ and a mono edge $e'$.
			\label{fig:Smallest5ArchMono1}}%
			{\includegraphics[page=1,scale=0.9]{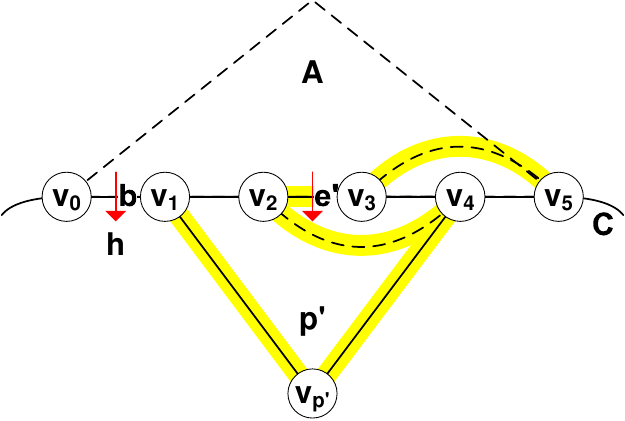}}
	\hspace{0.3cm}
	\subcaptionbox{$m_A = 6$, a mono edge $e'$, a non-mono edge $e$ and $v_3v_5 \in E(G)$.
			\label{fig:Smallest6ArchMono1}}%
			{\includegraphics[page=1,scale=0.9]{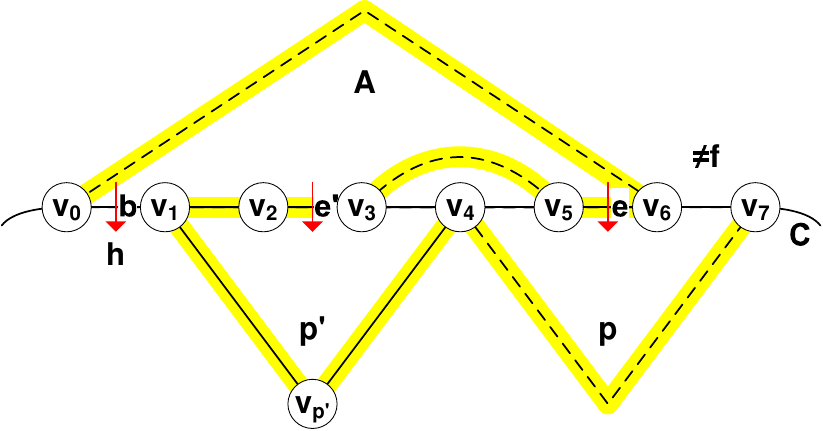}}
	\hspace{0.3cm}
	\subcaptionbox{$m_A = 6$, a mono edge $e'$, a non-mono edge $e$ and $v_3v_5 \notin E(G)$.
			\label{fig:Smallest6ArchMono2}}%
			{\includegraphics[page=1,scale=0.9]{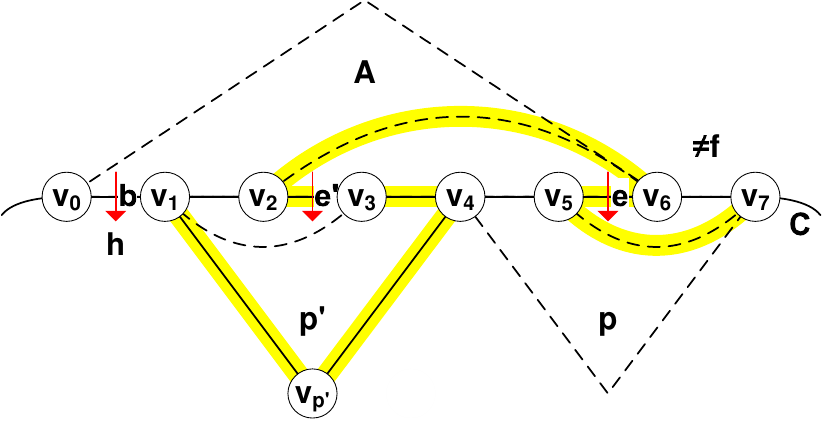}}
	\hspace{0.3cm}
	\caption{Case $m_A \geq 5$.
	}\label{fig:smallest5arches}
\end{figure}
		
		Hence, assume that $e'$ does not exist; then $e$ exists and $w(f) = 3$. In particular, $f$ has no opposite major face and $v_1v_2$ is not incident to $h$. Let $X \in \{C2,C7\}$. By the results about non-mono edges, then $v_3v_4$ is not incident to $p$. Hence, $v_2v_3$ is the middle $C$-edge of a minor 3-face, which is thick, as $h$ is thick. Then $v_2v_3$ is mono by $C3$ and Lemma~\ref{lem:conditionrestrictions} (for $C3$), which contradicts that $e'$ does not exist.
		
		Hence, $X \in \{C3,C4,C5,C6\}$. Since $v_1v_2$ is neither incident to $h$ nor to a major face nor to a minor thick 2-face, Lemma~\ref{lem:pedestalCost} implies $X = C3$. Then $e = v_{z-1}v_z$, as $e = v_{z-2}v_{z-1}$ contradicts by $C3$ that $e$ is non-mono. By definition of $C3$, then $v_2v_3$ is not incident to $p$. Since $v_2v_3$ is neither incident to $h$ nor to a minor thick 2-face, $v_2v_3$ is incident to a major face, which is a contradiction.
		
		\item[Case $m_A = 6$:]
		Then $w(f) = 3$, $e$ and $e'$ exist, $v_1v_2$ is not incident to $h$, and $f$ has no opposite major face (see Figure~\ref{fig:Smallest6ArchMono1}). Since $H$ has no minor 1-face, $e' = v_2v_3$, so that $v_1v_2$ is incident to $p'$. Then $X \notin \{C2,C7\}$, as otherwise $v_{z-2}v_{z-1}$ is either incident to a major face or to $p$, which contradicts $w(f) = 3$ due to the contribution of $\{v_{z-2}v_{z-1},\allowbreak v_{z-1}v_z,v_zv_{z+1}\}$.
		
		Hence, $X \in \{C3,C4,C5,C6\}$. By Lemma~\ref{lem:pedestalCost}, $X \in \{C3,C6\}$. By $C3$ (for $p'$), $v_2v_5 \notin E(G)$. If $X = C6$, $\deg_G(v_5) \geq 3$ and Lemma~\ref{lem:conditionrestrictions} imply $v_3v_5 \notin E(G)$ and $v_1v_5 \in E(G)$, which contradicts that $\{v_1,v_4\}$ is not a 2-separator of $G$. Hence, $X = C3$. By $C3$ (for $p$), $v_3v_6 \notin E(G)$. If $v_3v_5 \in E(G)$, $C$ is extendable by Figure~\ref{fig:Smallest6ArchMono1}, so assume otherwise. Since $\{v_1,v_4\}$ is not a 2-separator of $G$, we have $v_2v_6 \in E(G)$. Then $C$ is extendable by Figure~\ref{fig:Smallest6ArchMono2}.
	\end{description}
\end{description}
This completes the proof.
\end{proof}

In particular, we may choose $A$ in Lemma~\ref{lem:smallestkArch} as the proper arch of $f$.
}
This implies the following helpful corollary.

\begin{corollary}\label{cor:smallestkArch}
Every minor face $f$ that has a $C$-edge over which an opposite face of $f$ pulls weight by~$C2$ or $C7$ satisfies $w(f) \geq 2$ and, if $f$ is thick, $w(f) \geq 4$.
\end{corollary}

We now show that Inequalities~\eqref{eq:inequality1} and~\eqref{eq:inequality2} hold, which proves the Isolation Lemma.

\begin{lemma}\label{lem:suitableWeights}
Let $f$ be a face of $H$. Then \shortversion{$w(f) \geq 0$, $w(f) \geq 2$ if $f$ is thin and minor, and $w(f) \geq 4$ if $f$ is thick and minor.}
\longversion{
\begin{itemize}
	\item $w(f) \geq 0$ if $f$ is major,
	\item $w(f) \geq 2$ if $f$ is thin and minor, and
	\item $w(f) \geq 4$ if $f$ is thick and minor.
\end{itemize}
}
\end{lemma}
\begin{proof}
By Lemma~\ref{lem:atMost1OverAnyEdge}, any opposite face of $f$ pulls over any $C$-edge of $f$ weight at most one. Since the initial weight of such an edge for $f$ is one, we have $w(f) \geq 0$. In the remaining part of the proof, let $f$ be minor. Assume that $f$ has a $C$-edge $e'$ such that the $e'$-opposite face of $f$ pulls weight over $e'$ by Condition~C2 or~C7. Then the claim follows by Corollary~\ref{cor:smallestkArch}. We therefore assume throughout the proof that no opposite face of $f$ pulls weight over a $C$-edge of $f$ by Condition~$C2$ or $C7$.

Let $m_f = 2$. If $f$ is thick, Condition~$C2$ and Lemma~\ref{lem:atMost1OverAnyEdge} imply $w(f) = 4$. If $f$ is thin, assume to the contrary that $w(f) \leq 1$. Then $f$ has a $C$-edge $e$ such that \pulls{g}{e}{X} for the $e$-opposite face $g$ of $f$. By our assumptions, $X \in \{C3,C4,C5,C6\}$. Since $m_f = 2$, $X \notin \{C3,C4,C5\}$.
Thus, $X = C6$, which contradicts Lemma~\ref{lem:conditionrestrictions}.

\begin{figure}[!htb]
	\centering
	\captionbox{$m_f = 3$ when $e$ is an extremal $C$-edge of a 3-arch
			\label{fig:3FaceOverlapping}}[0.97\linewidth]
			{\includegraphics[page=1,scale=0.9]{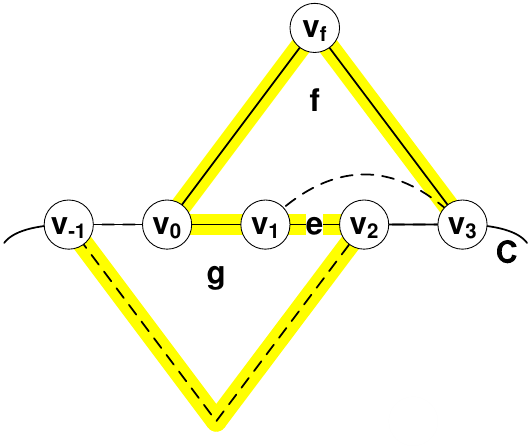}}
	\hspace{0.3cm}
\end{figure}

Let $m_f = 3$. Assume to the contrary that $f$ has a $C$-edge $b$ such that \pulls{h}{b}{X} for the $e$-opposite face $h$ of $f$. Then every $X \in \{C3,C4,C5,C6\}$ contradicts Lemma~\ref{lem:conditionrestrictions}. Hence, $w(f) \geq 3$. If $f$ is thin, this gives the claim, so let $f$ be thick. If $f$ has an opposite major face, $C1$ implies the claim $w(f) \geq 4$, so assume otherwise. Let $e$ be the middle $C$-edge of $f$ and let $g$ be the minor $e$-opposite face of $f$. Then $e$ is not an extremal $C$-edge of a 3-arch, as otherwise $C$ is extendable by Figure~\ref{fig:3FaceOverlapping}. If $m_g \geq 3$, we therefore have \pulls{f}{e}{C3}, which gives $w(f) \geq 4$. Otherwise, $m_g = 2$, since $H$ has no minor 1-face. By symmetry, say that $g$ has extremal $C$-vertices $v_0$ and $v_2$; then $\deg_G(v_1) \geq 3$ implies $v_1v_3 \in E(G)$, and Lemma~\ref{lem:conditionrestrictions} (for $C2$) implies that $g$ is thin. Then \pulls{f}{v_2v_3}{C2}, which gives $w(f) \geq 4$.

\shortversion{For $m_f = 4$, we defer the proof to the appendix due to space constraints.}
\longversion{
\begin{figure}[!htb]
	\centering
	\subcaptionbox{$e = v_0v_1$, $X = C3$ and $v_1v_3 \in E(G)$.
			\label{fig:4Face}}%
			{\includegraphics[page=1,scale=0.9]{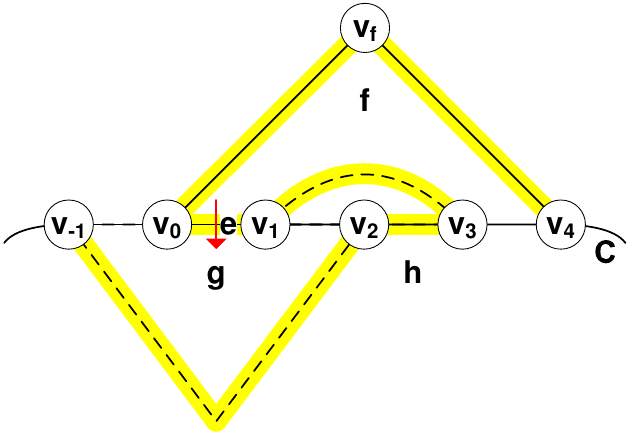}}
	\hspace{0.3cm}
	\subcaptionbox{$e = v_0v_1$ and $v_2v_3$ is an extremal $C$-edge of a 3-arch of $h$.
			\label{fig:4Face2}}%
			{\includegraphics[page=1,scale=0.9]{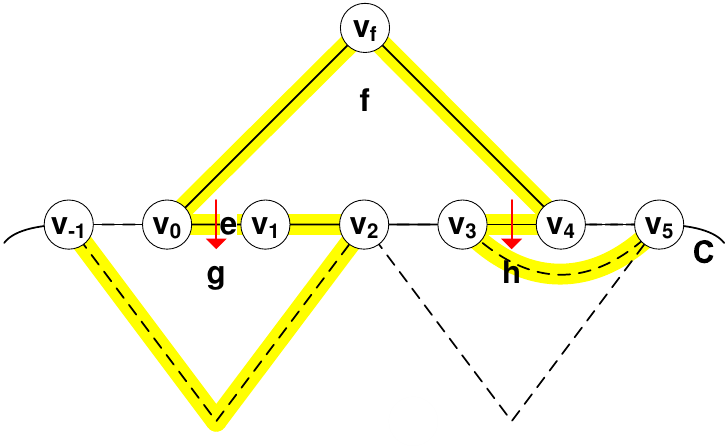}}
	\hspace{0.3cm}
	\caption{$m_f = 4$.
	}\label{fig:finallemma}
\end{figure}

Let $m_f = 4$. We may assume that $f$ has a $C$-edge $e$ such that \pulls{g}{e}{X} for the $e$-opposite face $g$ of $f$, as otherwise $w(f) \geq m_f = 4$. If $e = v_1v_2$ (or $e = v_2v_3$ by symmetry), Lemma~\ref{lem:conditionrestrictions} implies $X \notin \{C3,C4,C5\}$ because $m_f = 4$ and we have $X \neq C6$ by definition of~$C6$. We conclude that $\{v_1v_2,v_2v_3\}$ contributes weight at least two to $f$ and that $e$ is an extremal $C$-edge of $f$, say $e = v_0v_1$ by symmetry. If $f$ is thin, this gives the claim $w(f) \geq 2$, so let $f$ be thick. Since $e = v_0v_1$, $X \notin \{C4,C5\}$, as $v_{-1}v_0$ is not incident to $f$. Hence, $X \in \{C3,C6\}$, which implies in both cases that $v_2v_3$ is incident to a face $h \notin \{f,g\}$. By Lemma~\ref{lem:conditionrestrictions} (for $C3$ and~$C6$), $v_0v_3 \notin E(G)$. In addition, $v_1v_3 \notin E(G)$, as otherwise $X = C3$ implies that $C$ is extendable by Figure~\ref{fig:4Face} and $X = C6$ contradicts Lemma~\ref{lem:conditionrestrictions} (for $C6$).

If $v_2v_3$ is an extremal $C$-edge of a 3-arch, $v_0v_3 \notin E(G)$ and $v_1v_3 \notin E(G)$ imply $v_3v_5 \in E(G)$; then $C$ is extendable by Figure~\ref{fig:4Face2}, since $f$ is thick. Hence, $v_2v_3$ is not an extremal $C$-edge of a 3-arch. If an opposite face of $f$ pulls weight over $v_3v_4$ by some Condition~$Y$, $Y \in \{C3,C6\}$ by the same argument as for $X$; since $v_0v_3 \notin E(G)$ and $v_1v_3 \notin E(G)$, Lemma~\ref{lem:conditionrestrictions} (for $C6$) implies then $Y = C3$, so that $v_2v_3$ is an extremal $C$-edge of a 3-arch, which is impossible. Hence, no opposite face of $f$ pulls weight over $v_3v_4$.

If $h$ is major, we thus have $w(f) \geq 4$ by $C1$, so let $h$ be minor. Since $H$ has no minor 1-face and $v_2v_3$ contributes weight at least one to $f$, $m_h \geq 3$. If $v_1v_4 \in E(G)$, we have \pulls{f}{v_2v_3}{C3}, as $v_2v_3$ is not an extremal $C$-edge of a 3-arch, $h$ is minor, $m_h \geq 3$, $v_4v_5$ is not incident to $f$ and $v_1v_2$ is not incident to $h$. Since this gives the claim, let $v_1v_4 \notin E(G)$. If $X = C3$, then \pulls{f}{v_2v_3}{C6}, which gives the claim. If otherwise $X = C6$, then $\deg_G(v_1) \geq 3$, $v_1v_3 \notin E(G)$ and Lemma~\ref{lem:conditionrestrictions} (for $C6$) imply $v_1v_4 \in E(G)$, which is a contradiction.
}

\begin{figure}[!htb]
	\centering
	\subcaptionbox{$b = v_3v_4$, $X = C3$, $v_0v_4 \in E(G)$ and $v_{-1}v_1 \in E(G)$.
			\label{fig:5Face4a}}%
			{\includegraphics[page=1,scale=0.9]{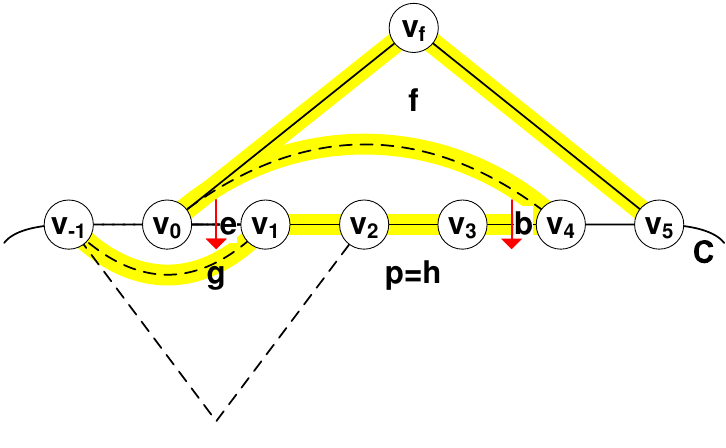}}
	\hspace{0.3cm}
	\subcaptionbox{$b = v_3v_4$, $X = C3$, $v_0v_4 \in E(G)$ and $v_1v_3 \in E(G)$.
			\label{fig:5Face4b}}%
			{\includegraphics[page=1,scale=0.9]{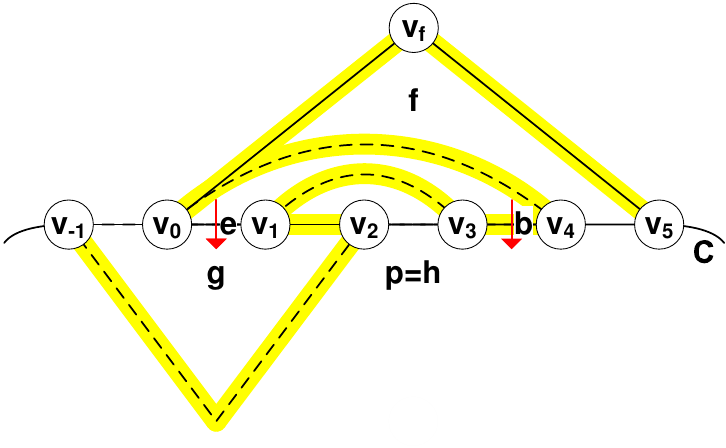}}
	\hspace{0.3cm}
	\subcaptionbox{$b = v_3v_4$, $X = C3$, $v_0v_4 \notin E(G)$.
			\label{fig:5Face4c}}%
			{\includegraphics[page=1,scale=0.9]{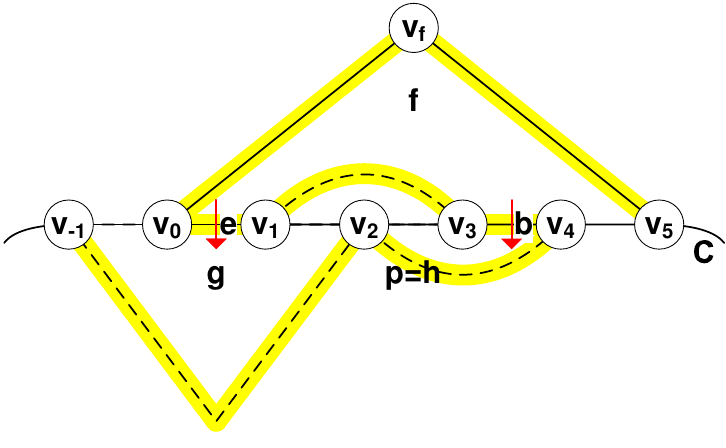}}
	\hspace{0.3cm}
	\caption{$m_f = 5$
	}\label{fig:finallemma2}
\end{figure}

Let $m_f \geq 5$. By Lemma~\ref{lem:notAdjacent}\ref{enum:notAdjacent}, $f$ has at most $\lfloor \frac{m_f+2}{3} \rfloor$ $C$-edges over which an opposite face of $f$ pulls weight. Hence, $w(f) \geq m_f - \lfloor \frac{m_f+2}{3} \rfloor = \lceil \frac{2}{3}(m_f-1) \rceil$. This gives the claim if $f$ is thin or $m_f \geq 6$, so let $f$ be thick, $m_f = 5$ and $w(f) = 3$.

Then $f$ has exactly two $C$-edges $e$ and $b$ such that \pulls{g}{e}{X} and \pulls{h}{b}{Y} for opposite faces $g$ and $h$ of $f$, and no major opposite face. By Lemma~\ref{lem:notAdjacent}\ref{enum:extremal} and~\ref{enum:notAdjacent}, $e$ or $b$ is an extremal $C$-edge of $f$, say $e = v_0v_1$ and $b \in \{v_3v_4,v_4v_5\}$ by symmetry. Since $v_{-1}v_0$ is not incident to $f$, $X \in \{C3,C6\}$, which implies in both cases that $v_2v_3$ is incident to a face $p \notin \{f,g\}$. Assume to the contrary that $b = v_4v_5$. Since $v_5v_6$ is not incident to $f$, $Y \in \{C3,C6\}$, which implies in both cases that $p \notin \{f,g,h\}$. Since $H$ has no minor 1-face, $p$ is major, which contradicts our assumption. We conclude that $b = v_3v_4$, so that $Y \in \{C3,C4,C5\}$ by definition of~$C6$.

By Lemma~\ref{lem:conditionrestrictions} for $X \in \{C3,C6\}$ and $Y \in \{C3,C4,C5\}$, $G$ does not contain any edge of $\{v_0v_3,v_1v_4,v_1v_5,v_2v_5\}$. Since $\deg_G(v_1) \geq 3$, $v_{-1}v_1 \in E(G)$ or $v_1v_3 \in E(G)$. As $X = C6$ implies $v_{-1}v_1 \notin E(G)$ and $v_1v_3 \notin E(G)$ by Lemma~\ref{lem:conditionrestrictions}, $X = C3$. If $v_0v_4 \in E(G)$, $C$ is extendable by Figure~\ref{fig:5Face4a} when $v_{-1}v_1 \in E(G)$ and by Figure~\ref{fig:5Face4b} otherwise. Hence, assume $v_0v_4 \notin E(G)$. By Lemma~\ref{lem:conditionrestrictions} for $Y \in \{C3,C4,C5\}$ and the fact that $\{v_2,v_5\}$ is not a 2-separator of $G$, a vertex of $\{v_0,v_1\}$ is adjacent to a vertex of $\{v_4,v_5\}$. By the previous results, $v_1v_3 \in E(G)$ and $v_2v_4 \in E(G)$. Then $C$ is extendable by Figure~\ref{fig:5Face4c}.
\end{proof}

\section{Algorithms}
We conclude this paper with algorithmic versions of the Isolation Lemma and of Theorem~\ref{thm:essential}.

\begin{theorem}\label{thm:algorithmicLemma}
Given an isolating cycle $C$ of length $c < \min\{\lfloor \frac{2}{3}(n+4) \rfloor,n\}$ of a polyhedral graph $G$ on $n$ vertices, a larger isolating cycle $C'$ of $G$ that satisfies $V(C) \subset V(C')$ and $|V(C')| \leq |V(C)| + 3 + n_5(G)$ can be computed in time $O(n)$.
\end{theorem}
\begin{proof}
We compute the graph $H$ and identify all minor faces of $H$, their $C$-edges, the information whether they are thick or thin, and all tunnels in linear time $O(n)$. If $H$ has a thick minor 1-face $f$, extending $C$ by the arch of $f$ in constant time gives the claim.
Otherwise, for every of the $2c = O(n)$ relevant face-edge pairs $(g,e)$, there is only a constant number of configurations in which $C$ is extended, according to the proof of the Isolation Lemma. We identify these cases and extend $C$ in total time $O(n)$.
\end{proof}

\begin{theorem}\label{thm:algorithmicEssential}
Given an essentially 4-connected planar graph $G$ on $n$ vertices, an isolating Tutte cycle of $G$ of length at least $\min\{\lfloor \frac{2}{3}(n+4) \rfloor,n\}$ can be computed in time $O(n^2)$.
\end{theorem}
\begin{proof}
If $n \leq 10$, $G$ is Hamiltonian\longversion{, as described in the proof of Theorem~\ref{thm:essential}}, so that we may compute even a Hamiltonian isolating cycle of $G$ in constant time. Hence, assume $n \geq 11$. For computing a first isolating Tutte cycle $C$ of $G$, we choose the start and end-edges of a Tutte path carefully as described in~\cite[Lemma~4(i)]{Fabrici2016}. This can be done in time $O(n^2)$ by using the algorithm from~\cite{Schmid2018} (we note that the faster algorithm by Biedl and Kindermann~\cite{Biedl2019} cannot be used here, as it does not allow to prescribe the start and end-edges). Then applying Theorem~\ref{thm:algorithmicLemma} iteratively to $C$ gives the claim.
\end{proof}

\paragraph{Acknowledgments.} The second author wants to thank Jochen Harant for introducing him to this interesting topic and helpful discussions.

\bibliographystyle{abbrv}
\bibliography{../../Jens}

\shortversion{
	\newpage
	\pagestyle{empty}
	\part*{\centering Appendix}
	\setboolean{@twoside}{false}
	\includepdf[pages=-,offset=0mm 0mm]{./Appendix.pdf}
}
\end{document}